%% file: neurips_2026_arxiv.tex
\documentclass{article}

\usepackage[main, final]{neurips_2026}

\usepackage[utf8]{inputenc} 
\usepackage[T1]{fontenc}    
\usepackage{hyperref}       
\usepackage{url}            
\usepackage{booktabs}       
\usepackage{amsfonts}       
\usepackage{nicefrac}       
\usepackage{microtype}      
\usepackage{xcolor}         
\usepackage{algorithm}
\usepackage{algpseudocode}
\usepackage{siunitx}
\usepackage{comment}
\usepackage{graphicx}
\usepackage{amsmath}
\usepackage{subcaption}\usepackage{multirow}

\title{Spectral-Aligned Pruning for Universal Error-Correcting Code Transformers}

\author{%
  \textbf{Sanghyeon Cho$^{1}$ \quad
  Taewoo Park$^{1}$ \quad
  Seong-Joon Park$^{2}$ \quad
  Dae-Young Yun$^{3}$} \\
  \textbf{Hee-Youl Kwak$^{3}$ \quad
  Sang-Hyo Kim$^{4}$ \quad
  Yongjune Kim$^{1}$\thanks{Corresponding author.}} \\
  \\
  $^{1}$Dept. of Electrical Engineering, POSTECH \\
  $^{2}$Institute of Artificial Intelligence, POSTECH \\
  $^{3}$Dept. of Electrical, Electronic and Computer Engineering, University of Ulsan \\
  $^{4}$Dept. of Electrical and Computer Engineering, Sungkyunkwan University \\
  \\
  \texttt{\{chosh320, parktaewoo, seongjoon, yongjune\}@postech.ac.kr} \\
  \texttt{\{yundy, hykwak\}@ulsan.ac.kr} \quad
  \texttt{iamshkim@skku.edu}
}

\begin{document}

\maketitle

\begin{abstract}

Universal channel decoders based on transformers-such as the Foundation Error Correction Code Transformer (FECCT)-achieve competitive decoding performance across diverse code families with a single shared backbone, optionally followed by code-specific finetuning. 
However, the high computational complexity and large parameter footprint of FECCT present substantial obstacles to practical deployment.
To address these challenges, we investigate structured pruning for FECCT and propose Spectral-Aligned Pruning (SAP), a structure-aware framework that enables cross-code reuse of structured pruning masks by leveraging the spectrum of the corresponding bipartite graph.
SAP is grounded in classical graph analysis of codes: the two algebraically largest adjacency eigenvalues provide compact spectral proxies for degree scale, expansion ratio, and minimum-distance lower bounds. 
These quantities are directly relevant to decoding performance: degree scale reflects how densely codeword bits and parity checks are connected; expansion ratio influences how information propagates across the bipartite graph; and minimum distance characterizes codeword separation.
Based on this connection, SAP uses these two leading eigenvalues as a lightweight code signature for pruning-mask retrieval. 
Empirically, this two-dimensional signature yields stable library selection equivalent to higher-dimensional spectral signatures in our evaluation.
After pruning, SAP uses parameter-efficient low-rank adaptation (LoRA) to recover decoding performance while storing only a small number of additional adapter parameters per code.
Experiments across diverse codes show that SAP achieves decoding performance comparable to dedicated per-code pruning, while enabling substantial reductions in computational cost and model memory footprint through kernel-level structured pruning.
Code is available at \url{https://github.com/iil-postech/sap}.

\end{abstract}

\section{Introduction}

Error-Correcting Codes (ECCs) are fundamental to reliable digital communication, enabling accurate recovery of information over noisy channels.
Over decades, coding theory has developed well-established code families and decoding algorithms by exploiting algebraic and graphical structures~\citep{MacWilliams1977The,Richardson2008modern}.
Each code family is usually accompanied by decoding algorithms that are optimized for its specific algebraic or graphical structure.

Transformers have demonstrated that self-attention can effectively model long-range dependencies through flexible token interactions~\citep{Vaswani2017attention}.
This capability has motivated neural ECC decoders such as the Error Correction Code Transformer (ECCT)~\citep{Choukroun2022error}, which incorporates bipartite graph structure into the decoding process through code-aware attention.
ECCT enables a single decoding \emph{algorithm} to operate across multiple code families; however, the resulting model architecture-and thus the number of parameters-depends on the code length and rate.
The Foundation Error Correction Code Transformer (FECCT) extends this direction by unifying not only the decoding algorithm but also the model \emph{architecture and parameter set}~\citep{Choukroun2024foundation}.
In particular, FECCT uses a single decoder architecture with a fixed parameter set shared across different code families and code parameters, while allowing the parameter values to be retrained for individual codes.

Despite these advances, transformer-based decoders remain expensive in terms of computational cost and model memory footprint, limiting their practicality in resource-constrained receivers.
To mitigate the computational and memory overheads of transformer-based ECC decoders, several approaches have been explored.
Architectural approaches, such as Cross-Attention Message-Passing Transformer (CrossMPT)~\citep{Park2025crossmpt} and Efficient Message-Passing Transformer (EfficientMPT)~\citep{Park2026Efficient}, redesign the attention mechanism to improve efficiency.
In addition, quantization-based approaches have been proposed, including Accelerating Error Correction Code Transformers (AECCT)~\citep{Levy2024accelerating}, which adopt ternary weight quantization.
However, structured pruning for transformer-based ECC decoders has not been systematically studied.

In this paper, we investigate \emph{structured pruning} for universal transformer-based decoders.
In transformer architectures, structured pruning removes attention heads and feed-forward network (FFN) channels, yielding practical kernel-level speedups~\citep{Michel2019are,Voita2019analyzing}.
A pruning mask specifies which structural units (e.g., attention heads and FFN channels) are retained or removed, and is typically derived from importance scores computed on a calibration set, incurring significant computational overhead~\citep{Zafrir2021prune,Kwon2022fast,Cheng2024survey}.
In the context of ECC decoders, it is important to determine pruning masks that preserve decoding performance.
A straightforward approach is to derive a separate pruning mask for each code; however, this becomes impractical for universal decoders that must support a large and diverse set of codes.

To address this limitation, we formulate the \emph{multi-code structured pruning} problem for universal transformer-based ECC decoders.
This naturally raises a key question:
\emph{When can structured pruning masks be shared across different codes while maintaining decoding performance?}
Answering this question requires comparing codes in a meaningful way. Codes vary in block length, rate, and parity-check structure, making direct comparison difficult.
Efficient mask reuse therefore requires a code descriptor that quantifies structural similarity across heterogeneous codes and is grounded in code-theoretic properties relevant to decoding behavior.

To answer this question, we propose \emph{Spectral-Aligned Pruning (SAP)}, a structure-aware framework that provides a principled criterion for reusing structured pruning masks across different codes.
SAP enables cross-code reuse by aligning codes according to the leading adjacency eigenvalues of their parity-check bipartite graphs.
This choice is grounded in classical graph analysis of codes: the two algebraically largest adjacency eigenvalues $\lambda_1, \lambda_2$ provide compact spectral proxies for decoding-relevant structural properties of the bipartite graph.
Specifically, $\lambda_1$ reflects the degree scale (the connection density between codeword bits and parity checks)~\citep{Brouwer2011spectra}, the spectral gap $\lambda_1 - \lambda_2$ characterizes expansion ratio and influences how information propagates across the bipartite graph~\citep{Sipser1996Expander,Hoory2006expander}, and $\lambda_1, \lambda_2$ together appear in classical minimum-distance analysis~\citep{Tanner2001minimum,Shin2005Generalization}.
Codes with similar leading eigenvalues thus share similar fundamental properties relevant to decoding behavior, motivating SAP's use of just two eigenvalues as a lightweight, theoretically-grounded code signature.
We empirically verify that this two-dimensional signature selects the same pruning masks as higher-dimensional spectral signatures in our evaluated code set.

Given a target code, SAP computes the two leading adjacency eigenvalues and retrieves the nearest entry from a pruning mask library using a spectral distance over this two-dimensional signature.
If the spectral similarity score to the nearest neighbor exceeds a predefined threshold, SAP reuses the associated mask to prune FECCT for the target code.
Otherwise, SAP derives a new structured pruning mask using code-conditioned importance estimation and updates the library by inserting the resulting spectral signature and pruning mask pair.
By using spectral proximity as the reuse criterion and expanding the library only when necessary, SAP amortizes the pruning mask search cost while avoiding unreliable sharing across structurally dissimilar bipartite graphs.

To efficiently recover decoding performance after pruning, we adopt a lightweight and parameter-efficient recovery step using \emph{low-rank adaptation (LoRA)} adapters on top of the pruned backbone.
We freeze the parameters of the pruned backbone and fine-tune only the code-specific low-rank adapter parameters.
This design enables storage-efficient multi-code deployment. SAP stores a single unpruned FECCT backbone, one binary pruning mask per library entry, and one LoRA adapter per code, and materializes the pruned model for a target code on demand. This substantially reduces storage overhead compared to storing a fully fine-tuned model for each code.

\textbf{Contributions.}

\textbf{(C1) Multi-code structured pruning problem.}
We formulate the \emph{multi-code structured pruning} problem for universal transformer-based ECC decoders, where a single shared backbone must support heterogeneous codes with different block lengths, rates, and parity-check structures.
Unlike conventional structured pruning settings that typically prune a model for a fixed task or input structure, this setting requires deciding when a pruning mask derived for one code can be safely reused for another code.

\textbf{(C2) Spectral-aligned pruning.}
We propose Spectral-Aligned Pruning (SAP), a structured pruning framework that enables cross-code pruning-mask reuse through thresholded nearest-neighbor retrieval in a spectral mask library. 
Motivated by classical spectral graph analysis of codes, SAP uses the two largest adjacency eigenvalues of the parity-check bipartite graph as a lightweight code signature. We empirically verify that this two-dimensional signature yields the same library-mask selections as higher-dimensional spectral signatures in our evaluated code set and strongly correlates with cross-code pruning-mask overlap.

\textbf{(C3) Efficient multi-code deployment.}
SAP combines structured pruning with parameter-efficient LoRA-based recovery, yielding shared pruning masks with compact code-specific adapters.
Across BCH, LDPC, Polar, and 5G NR LDPC codes, including a longer LDPC $(512,474)$ code, SAP achieves decoding performance comparable to dedicated per-code pruning while substantially reducing FLOPs, the memory footprint of the pruned model, and per-code mask-derivation cost.

\section{Background}

\subsection{Error-Correcting Codes}\label{sec:ecc}

We consider an $(n,k)$ binary linear block code with rate $k/n$.
A code is specified by a generator matrix $G\in\{0,1\}^{k\times n}$ and a parity-check matrix $H\in\{0,1\}^{(n-k)\times n}$ satisfying $GH^\top=0$ over $\mathrm{GF}(2)$.
Encoding maps a message $m\in\{0,1\}^k$ to a codeword $x=mG\in\{0,1\}^n$ such that $Hx^\top = 0$ over $\mathrm{GF}(2)$.
We adopt the standard binary phase-shift keying (BPSK) over an additive white Gaussian noise (AWGN) model as in~\citep{Choukroun2022error,Choukroun2024foundation}. 
Bits are modulated as $x_s = 1-2x\in\{-1,+1\}^n$ and transmitted over the AWGN channel with output $y=x_s+z$, where $z\sim\mathcal{N}(0,\sigma^2 I)$. 
Decoding aims to recover $x$ from $y$, and performance is evaluated in terms of bit error rate (BER) versus $E_b/N_0$.

\subsection{Transformer-based ECC Decoders}\label{sec:ecct}

Transformer architectures have been applied to the problem of ECC decoding.
ECCT incorporates transformer inference into the decoding process by operating on a codeword-invariant input representation and injecting code structure through PCM-derived masked self-attention~\citep{Choukroun2022error}.
FECCT extends this paradigm to a single universal decoder by sharing model parameters across multiple code families and block lengths, while modulating attention based on distances in the bipartite graph. 
In FECCT, decoding performance is typically recovered through retraining, which involves updating the full set of model parameters~\citep{Choukroun2024foundation}.
Despite their flexibility, attention heads and FFN blocks in transformer-based ECC decoders incur substantial computational cost and model memory footprint, necessitating efficiency improvements.

Several recent works address these computational and memory bottlenecks by redesigning attention architectures to better exploit code structure~\citep{Park2025crossmpt, Park2026Efficient}, as well as by adopting weight quantization to reduce model footprint~\citep{Levy2024accelerating}.
In contrast to architectural redesigns or quantization-based approaches, we investigate \emph{structured pruning} for transformer-based ECC decoders, removing attention heads and FFN channels to reduce computational cost and model memory footprint while preserving universal decoding capability.

\subsection{Structured Pruning of Transformers}\label{sec:pruning}

We build on post-training structured pruning methods for transformers, which remove coarse-grained structural units such as attention heads and FFN channels to obtain practical kernel-level speedups.
A pruning mask specifies which units are retained or removed under a given compute budget and can be derived using a range of importance criteria, including magnitude- or activation-based heuristics, gradient- or Taylor-based approximations, and second-order sensitivity estimates (e.g., Fisher- or Hessian-based scores) computed on a small calibration set~\citep{Zafrir2021prune, Kwon2022fast, Cheng2024survey}.
In our work, we adopt a Fisher-based importance estimator~\citep{Kwon2022fast} as an effective primitive for deriving structured pruning masks from transformer-based ECC decoders.

\begin{figure*}[!t]
\vskip 0.1in
\begin{center}
\includegraphics[width=\textwidth]{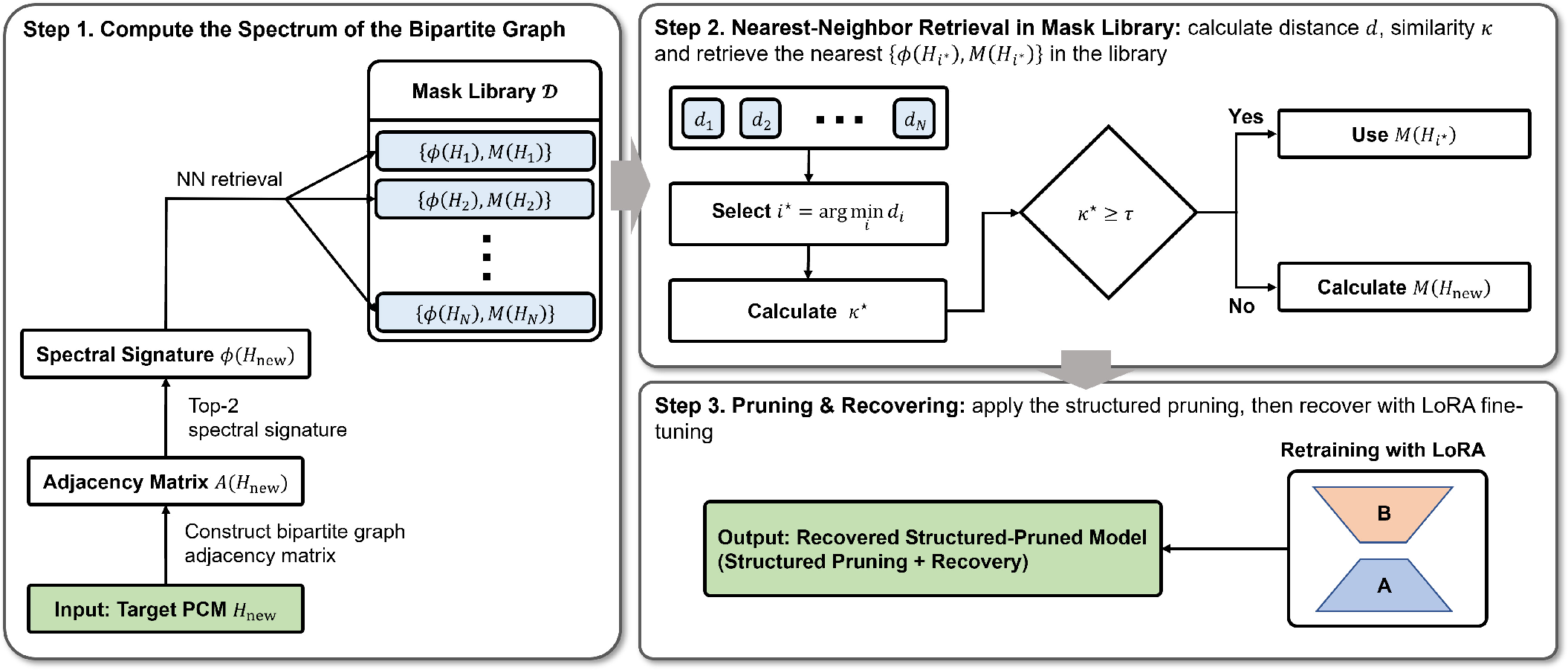}
\caption{Overview of SAP. For a target PCM, SAP computes a two-eigenvalue spectral signature, retrieves or derives a pruning mask through thresholded library matching, applies the selected mask to the common FECCT backbone, and performs code-specific LoRA recovery.} \label{fig:pipeline}
\end{center}
\vskip -0.3in
\end{figure*}

\section{SAP: Spectral-Aligned Pruning}\label{sec:method}

To effectively address the \emph{multi-code structured pruning} problem, we propose the SAP framework (see Figure~\ref{fig:pipeline}). 
Since deriving a structured pruning mask via code-conditioned importance estimation is costly to repeat for each new code, SAP amortizes this overhead by maintaining a library of previously derived code-mask pairs and reusing a stored mask when the target code is sufficiently similar in terms of its spectral signature.
Specifically, SAP computes a spectral signature for $H_{\text{new}}$ (Section~\ref{sec:spectrum}), performs nearest-neighbor retrieval followed by a threshold-based reuse decision (Section~\ref{sec:library}), and then applies structured head/FFN pruning with a subsequent parameter-efficient recovery step (Section~\ref{sec:recovery}).

\subsection{Bipartite Graph Spectrum}\label{sec:spectrum}

SAP aims to avoid deriving a new structured pruning mask for every target code by reusing a pruning mask obtained from a previously processed code.
To enable such reuse, we require a comparison criterion that reflects the code-dependent structure used by FECCT while remaining applicable to codes with different block lengths, rates, and parity-check structures.
We use the adjacency spectrum of the bipartite graph induced by the parity-check matrix as this criterion.

Given a parity-check matrix \(H \in \{0,1\}^{(n-k)\times n}\), the adjacency matrix of the corresponding bipartite graph is given by
\[
A(H)=
\begin{bmatrix}
0 & H^\top \\
H & 0
\end{bmatrix}
\in \{0,1\}^{(2n-k)\times(2n-k)}.
\]
This matrix \(A(H)\) captures the graph structure from which the code-dependent attention connectivity in FECCT is constructed.

Since different codes can have different block lengths and rates, the size of \(A(H)\) varies across codes, making direct matrix-level comparison unsuitable as a common criterion.
SAP therefore constructs a fixed-dimensional spectral signature from the adjacency spectrum of \(A(H)\).
The adjacency spectrum compactly summarizes structural properties of the bipartite graph, and, as shown in Section~\ref{sec:analysis-spectrum-mask}, this spectral descriptor exhibits a strong correlation with structured pruning-mask similarity.

Denote by \(\{\lambda_i\}_{i=1}^{2n-k}\) the eigenvalues of \(A(H)\), ordered in descending algebraic value:
\begin{equation}
\lambda_1 \ge \lambda_2 \ge \cdots \ge \lambda_{2n-k}.
\label{eq:eigenvalue}
\end{equation}
The two algebraically largest eigenvalues \(\lambda_1, \lambda_2\) capture decoding-relevant structural properties of the bipartite graph. 
Specifically, \(\lambda_1\) reflects the degree scale, characterizing the connection density between codeword bits and parity checks~\citep{Brouwer2011spectra}, while the spectral gap \(\lambda_1 - \lambda_2\) characterizes graph expansion that influences how information propagates across the bipartite graph~\citep{Sipser1996Expander,Hoory2006expander}.
In addition, both eigenvalues appear in classical minimum-distance analysis of bipartite code graphs~\citep{Tanner2001minimum,Shin2005Generalization}.
A more detailed discussion of these graph-spectral properties is provided in Appendix~\ref{app:spectrum_ecc}.

Based on these observations, we define the spectral signature as
\begin{equation}
\phi(H)=[\lambda_1,\lambda_2]\in\mathbb{R}^2.
\label{eq:sig}
\end{equation}

For two codes with PCMs \(H_A\) and \(H_B\), we define the spectral distance as
\begin{equation}
d(\phi(H_A),\phi(H_B))
=
\|\phi(H_A)-\phi(H_B)\|_2.
\label{eq:signature}
\end{equation}
The corresponding spectral similarity score is then defined
as
\begin{equation}
\kappa(H_A,H_B)=\exp\!\left(-\, \beta \, d(\phi(H_A),\phi(H_B))\right)\in(0,1],
\label{eq:spec_sim}
\end{equation}
where we fix \(\beta=0.1\) throughout all experiments.
Since $\kappa$ is a monotone function, the choice of $\beta$ does not affect the relative ordering of codes by spectral similarity.

\subsection{Pruning Mask Library and Spectral Retrieval}\label{sec:library}

SAP maintains a pruning mask library to enable mask reuse and avoid repeated pruning mask derivation across different codes.
For each code, the library stores the spectral signature of its bipartite graph and the corresponding structured pruning mask for the pretrained FECCT.

Given a code specified by $H$, the library stores an entry 
$\mathcal{D}(H) \;=\; \left(\phi(H),\, M(H)\right)$, 
where $\phi(H)\in\mathbb{R}^2$ is the spectral signature defined in~\eqref{eq:sig}, and $M(H)$ denotes the structured pruning mask given by
$M(H) \;=\; \{ (m^{(\ell)}_{\text{head}}(H),\, m^{(\ell)}_{\text{FFN}}(H)) \}_{\ell=1}^{L}$.
Here, $m^{(\ell)}_{\text{head}}(H)$ and $m^{(\ell)}_{\text{FFN}}(H)$ denote the binary mask vectors corresponding to the attention heads and FFN channels in the $\ell$-th layer, respectively. 
For each mask, an element value of $0$ indicates that the corresponding attention head or FFN channel is pruned, while a value of $1$ indicates that it is retained.
When the dependence on $H$ is clear from context, we omit it for notational simplicity.

We initialize the library with $N$ reference codes, storing their spectral signatures $\phi(H)$ and pruning masks $M(H)$ derived from the pretrained FECCT, i.e., $\{\mathcal{D}(H_i)=(\phi(H_i),M(H_i))\}_{i=1}^{N}$.
Given a target code specified by $H_{\text{new}}$, we compute $\phi(H_{\text{new}})$ using~\eqref{eq:sig}. 
We then identify the nearest reference code based on the spectral similarity score defined in~\eqref{eq:spec_sim} as follows:
\begin{equation}
\begin{aligned}
i^\star = \operatorname*{arg\,min}_{i\in\{1,\dots,N\}} \; d\left( \phi(H_{\text{new}}), \phi(H_i)\right), \qquad 
\kappa^\star = \kappa(H_{\text{new}},H_{i^\star}).
\end{aligned}
\label{eq:nn_retrieval}
\end{equation} 

Based on the spectral similarity score $\kappa^\star$, SAP applies the following threshold-based reuse rule:
\begin{equation}
\widehat{M}(H_{\text{new}}) \;=\;
\begin{cases}
M(H_{i^\star}), & \text{if } \kappa^\star \ge \tau,\\
M(H_{\text{new}}), & \text{otherwise},
\end{cases}
\label{eq:reuse_rule}
\end{equation}
where $\tau$ is a similarity threshold. 
Here, $M(H_{i^\star})$ denotes the pruning mask retrieved from the library, whereas $M(H_{\text{new}})$ denotes a newly computed code-specific pruning mask obtained via the code-conditioned importance estimation procedure. 
Specifically, if $\kappa^\star \ge \tau$, SAP reuses the retrieved pruning mask $M(H_{i^\star})$; otherwise, SAP derives a new code-specific mask and updates the library by inserting the new entry:
\begin{equation}
\mathcal{D} \leftarrow \mathcal{D} \cup \{(\phi(H_{\text{new}}), M(H_{\text{new}}))\}
\label{eq:library_update}
\end{equation}
This mechanism restricts mask reuse to spectrally aligned codes while progressively incorporating codes that are dissimilar to existing entries into the library.
Algorithm~\ref{alg:sap} in Appendix \ref{app:MS_algorithm} summarizes this overall procedure.

\subsection{Parameter-Efficient Retraining}\label{sec:recovery}

To recover decoding performance degraded by structured pruning without maintaining a separately retrained model for each code, we perform per-code recovery using LoRA adapters on top of the pruned backbone~\citep{Hu2022lora,Zimmer2023perp}.
This recovery step follows the practical adaptation protocol of FECCT, where achieving strong decoding performance for a target code typically requires code-specific fine-tuning of the shared universal backbone~\citep{Choukroun2024foundation}. 
Unlike full code-specific fine-tuning, SAP keeps the pruned backbone frozen, and trains only lightweight code-specific LoRA parameters for adaptation.
This design substantially reduces the memory overhead compared to saving a fully retrained model for each code.
We train these adapters using the binary cross-entropy (BCE) loss, optionally augmented with a lightweight knowledge distillation (KD) from the pretrained (unpruned) FECCT, following prior work on pruning with distillation~\citep{Muralidharan2024compact}.

For a received sample $y\in\mathbb{R}^n$, let $f_{\theta_t}(y), f_{\theta_s}(y)\in\mathbb{R}^n$ be the outputs of the teacher (unpruned) and student (pruned) models, respectively.
We convert the logits into bitwise posterior probabilities via $q_{t,j}=\sigma\!\big(\mathrm{sign}(y_j)\,f_{\theta_t}(y)_j\big),\ 
q_{s,j}=\sigma\!\big(\mathrm{sign}(y_j)\,f_{\theta_s}(y)_j\big)$,
and define the KD loss as the average Kullback-Leibler (KL) divergence between bitwise posteriors. 

The overall recovery objective is
\begin{equation}
\mathcal{L}=\mathcal{L}_{\mathrm{BCE}}+\gamma\,\mathcal{L}_{\mathrm{KD}},
\label{eq:loss}
\end{equation}
where $\gamma$ is a weighting hyperparameter.

We implement this recovery by training LoRA adapters on top of the pruned backbone while keeping all backbone parameters frozen~\citep{Hu2022lora}.
For a weight matrix $W\in\mathbb{R}^{d_{\mathrm{out}}\times d_{\mathrm{in}}}$, LoRA parameterizes a low-rank update as
$W' \;=\; W + \Delta W,\ \Delta W \;=\; BA$,
where $A\in\mathbb{R}^{r\times d_{\mathrm{in}}}$ and $B\in\mathbb{R}^{d_{\mathrm{out}}\times r}$ with rank $r\ll \min(d_{\mathrm{in}},d_{\mathrm{out}})$. 
Recovery optimizes only the adapter parameters under the objective in~\eqref{eq:loss}. 
After retraining, the adapters can be merged into the corresponding weights.

Because recovery updates only low-rank adapters, SAP stores a single unpruned FECCT backbone, one binary pruning mask per library entry, and one compact LoRA adapter per code. 
For a target code, the pruned model is materialized on demand by applying the mask of the retrieved library entry to the backbone and attaching the code-specific adapter. 
Codes assigned to the same library entry share the mask but keep separate adapters.
This substantially reduces per-code storage compared to keeping a separate fully retrained model for each code, while restoring pruning-induced decoding performance degradation in practice.

\begin{table*}[t]
\caption{Efficiency summary under structured pruning (\SI{40}{\percent} FLOPs-based ratio).
We report FLOPs and parameter counts for the full FECCT and the structurally pruned model (after physically removing pruned attention heads and FFN channels), along with the \emph{per-code} LoRA adapter size.
FLOPs reduction (Red.) (\%) is computed as $100\times(\text{Full}-\text{Pruned})/\text{Full}$.
}
\label{tab:efficiency}
\vspace{-3pt}
\begin{center}
\resizebox{0.9\linewidth}{!}
{
\begin{tabular}{c ccc cccc}
\toprule
\multirow{2}{*}{Codes} &
\multicolumn{3}{c}{FLOPs} &
\multicolumn{4}{c}{Parameters} \\
\cmidrule(lr){2-4}\cmidrule(lr){5-8}
& Full (M) & Pruned (M) & Red. (\%) &
  Full (M) & Pruned (M) & LoRA (M) & Ratio (LoRA / Full) \\
\midrule
BCH $(31,16)$    & 115.03 &  69.00 & \multirow{6}{*}{40.00} & \multirow{6}{*}{1.22} & 0.73 & 0.090 & 7.38\% \\
BCH $(63,45)$    & 211.26 & 126.74 &                       &                       & 0.72 & 0.089 & 7.30\% \\
LDPC $(96,64)$   & 403.44 & 242.05 &                       &                       & 0.70 & 0.088 & 7.21\% \\
LDPC $(121,70)$  & 510.39 & 306.20 &                       &                       & 0.68 & 0.087 & 7.13\% \\
Polar $(64,32)$  & 254.80 & 152.86 &                       &                       & 0.70 & 0.089 & 7.39\% \\
Polar $(128,64)$ & 566.23 & 339.74 &                       &                       & 0.67 & 0.086 & 7.05\% \\
\bottomrule
\end{tabular}}
\end{center}
\vskip -0.22in
\end{table*}

\section{Experiments} \label{sec:experiments}

In this section, we evaluate the proposed \emph{SAP} framework on FECCT, as it supports a unified transformer decoder architecture with a consistent parameter set across multiple code families, block lengths, and rates, which aligns with the design assumptions of SAP.

\textbf{Pretraining.} 
The FECCT backbone is pretrained on BCH, LDPC, and polar codes, using the 12 codes summarized in  Appendix~\ref{app:pretrain_codes}.
Pretraining is conducted on the all-zero codewords for $4,000$ epochs with a batch size of $256$, where $E_b/N_0$ is uniformly sampled between \SI{2}{\decibel} and \SI{7}{\decibel}.
We use the Adam optimizer and a cosine learning-rate schedule, decaying from $1\times 10^{-4}$ to $1\times 10^{-6}$.
Full implementation details are summarized in Appendix~\ref{app:implementation_details}.

\textbf{SAP library.}
We construct a pruning mask library using six representative ECCs commonly adopted in ECCT-based decoder evaluations, selected to span diverse code families and parameters: BCH $(31,16)$ and $(63,51)$, LDPC $(96,64)$ and $(121,60)$, and polar $(64,48)$ and $(128,86)$ codes~\citep{Choukroun2022error,Choukroun2024foundation}.
We use $\tau=0.5$ as the reuse threshold (Appendix~\ref{app:tau_threshold}).

\textbf{Retraining.}
For a target code, we perform LoRA-based recovery after structured pruning with rank $r=8$ and scaling factor $\alpha=16$.
LoRA adapters are applied to all attention projection matrices ($W_Q$, $W_K$, $W_V$, and $W_O$) in the transformer.
The pruned model is retrained for 100 epochs using both the BCE loss and the KD loss, with the unpruned model as the teacher; in this case, we set $\gamma=1$ in~\eqref{eq:loss}.
All experiments were run on NVIDIA RTX A6000 GPUs.

\subsection{Efficiency Gains: FLOPs and Parameter Counts} \label{sec:efficiency}

We evaluate the efficiency benefits of our approach along three dimensions: computational cost (FLOPs), the parameter count of the pruned model, and the additional \emph{per-code} parameters required for LoRA-based recovery.
Unless otherwise stated, we use a \SI{40}{\percent} pruning ratio as the default pruning ratio for the efficiency measurements.

Table~\ref{tab:efficiency} shows that structured pruning yields consistent efficiency gains in FLOPs and memory footprint across all evaluated codes.
Since we apply a fixed FLOPs-based pruning ratio of \SI{40}{\percent}~\citep{Kwon2022fast}, the FLOPs reduction is \SI{40}{\percent} for every code by design. 
Moreover, physically removing pruned attention heads and FFN channels correspondingly reduces the parameter footprint of the pruned model. 
To support many codes without storing a fully retrained model for each one, we adopt LoRA and store only compact, per-code adapters, while codes in the same library entry share a pruning mask applied to a single FECCT backbone.
Table~\ref{tab:efficiency} quantifies the resulting storage overhead: the per-code LoRA parameters account for only \SIrange{7}{7.4}{\percent} of the full FECCT parameter count.
This enables storage-efficient multi-code deployment, with decoding performance evaluated in the following section.
Additional wall-clock mask-derivation costs, memory measurements, and end-to-end inference latency are reported in Appendices~\ref{app:mask_cost} and~\ref{app:memory_footprint}.

\begin{table*}[t]
\caption{Decoding performance at \(E_b/N_0 \in \{4,5,6\}\) \SI{}{\decibel}. 
Entries are \(-\ln(\mathrm{BER})\) (higher is better) as in~\citep{Choukroun2022error,Choukroun2024foundation}. 
Parentheses indicate the performance gap \(\Delta = \text{SAP} - \text{Dedicated}\). 
Across both seen and unseen codes, SAP achieves decoding performance comparable to dedicated per-code pruning, showing no material degradation under mask reuse.}
\label{tab:decoding_performance}
\vspace{-5pt}
\begin{center}
\resizebox{0.98\linewidth}{!}
{
\begin{tabular}{l c c ccc ccc}
\toprule
\multirow{2}{*}{\textbf{Category}} & \multirow{2}{*}{\textbf{Target Codes}} & \multirow{2}{*}{\textbf{Library Ref. Codes}} 
& \multicolumn{3}{c}{\textbf{Dedicated Masks}} 
& \multicolumn{3}{c}{\textbf{SAP Masks}} \\
\cmidrule(lr){4-6}\cmidrule(lr){7-9}
& & & \SI{4}{\decibel} & \SI{5}{\decibel} & \SI{6}{\decibel} & \SI{4}{\decibel} & \SI{5}{\decibel} & \SI{6}{\decibel} \\
\midrule
\multirow{6}{*}{Seen Codes}
& BCH $(63,36)$    & BCH $(63,51)$    & 4.53 & 6.30 & 8.92 & 4.51 ($\mathbf{-0.02}$) & 6.30 ($\mathbf{+0.00}$) & 9.00 ($\mathbf{+0.08}$) \\
& BCH $(63,45)$    & BCH $(63,51)$    & 5.16 & 7.33 & 10.32 & 5.18 ($\mathbf{+0.02}$) & 7.30 ($\mathbf{-0.03}$) & 10.16 ($-0.16$) \\
& LDPC $(49,24)$   & LDPC $(121,60)$  & 6.16 & 8.83 & 12.67 & 6.15 ($\mathbf{-0.01}$) & 8.80 ($\mathbf{-0.03}$) & 12.69 ($\mathbf{+0.02}$) \\
& LDPC $(121,70)$  & LDPC $(121,60)$  & 6.43 & 10.20 & 15.80 & 6.43 ($\mathbf{+0.00}$) & 10.12 ($\mathbf{-0.08}$) & 15.65 ($\mathbf{-0.15}$) \\
& Polar $(64,32)$  & Polar $(64,48)$  & 6.10 & 8.37 & 11.52 & 6.16 ($\mathbf{+0.06}$) & 8.35 ($\mathbf{-0.02}$) & 11.48 ($\mathbf{-0.04}$) \\
& Polar $(128,64)$ & Polar $(128,86)$ & 5.21 & 7.48 & 10.61 & 5.20 ($\mathbf{-0.01}$) & 7.47 ($\mathbf{-0.01}$) & 10.47 ($\mathbf{-0.14}$) \\
\midrule
\multirow{7}{*}{Unseen Codes}
& BCH $(31,11)$    & BCH $(31,16)$    & 4.83 & 6.37 & 8.53 & 4.84 ($\mathbf{+0.01}$) & 6.37 ($\mathbf{+0.00}$) & 8.47 ($\mathbf{-0.06}$) \\
& BCH $(31,21)$    & BCH $(31,16)$    & 6.19 & 8.36 & 11.20 & 6.20 ($\mathbf{+0.01}$) & 8.26 ($\mathbf{-0.10}$) & 11.20 ($\mathbf{+0.00}$) \\
& LDPC $(32,16)$   & BCH $(31,16)$    & 5.12 & 6.89 & 9.15 & 5.15 ($\mathbf{+0.03}$) & 6.93 ($\mathbf{+0.04}$) & 9.09 ($\mathbf{-0.06}$) \\
& LDPC $(64,48)$   & LDPC $(96,64)$   & 7.40 & 10.07 & 13.27 & 7.40 ($\mathbf{+0.00}$) & 10.04 ($\mathbf{-0.03}$) & 13.36 ($\mathbf{+0.09}$) \\
& LDPC $(96,72)$   & LDPC $(96,64)$   & 7.31 & 10.32 & 13.85 & 7.33 ($\mathbf{+0.02}$) & 10.23 ($\mathbf{-0.09}$) & 13.72 ($\mathbf{-0.13}$) \\
& Polar $(64,43)$  & Polar $(64,48)$  & 6.50 & 8.96 & 11.64 & 6.48 ($\mathbf{-0.02}$) & 8.90 ($\mathbf{-0.06}$) & 11.66 ($\mathbf{+0.02}$) \\
& Polar $(128,96)$ & Polar $(128,86)$ & 6.04 & 8.75 & 12.06 & 6.02 ($\mathbf{-0.02}$) & 8.72 ($\mathbf{-0.03}$) & 12.26 ($\mathbf{+0.20}$) \\
\bottomrule
\end{tabular}}
\end{center}
\vskip -0.3in
\end{table*}

\begin{figure*}[t]
\vskip 0.2in
\begin{center}
\begin{subfigure}[b]{.30\textwidth}
    \centering
    \includegraphics[width=\textwidth]{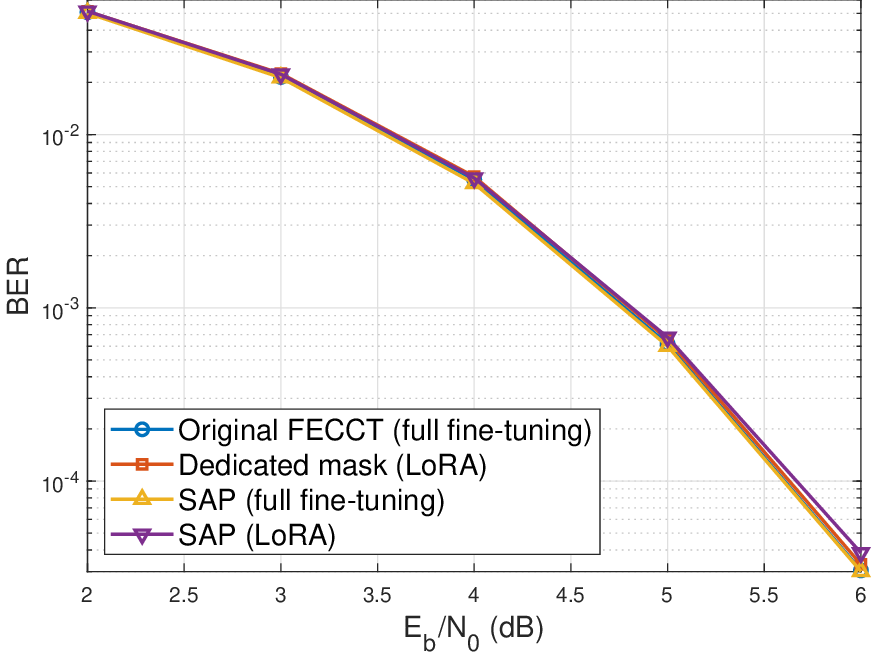}
    \caption{BCH $(63,45)$}
    \label{fig_63_45}
\end{subfigure}
\hfill
\begin{subfigure}[b]{.30\textwidth}
    \centering
    \includegraphics[width=\textwidth]{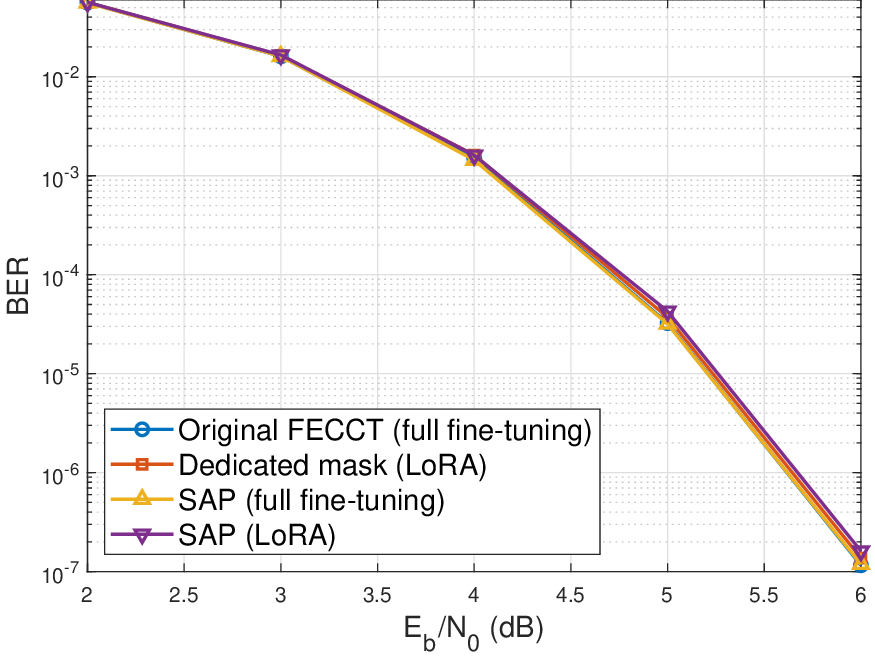}
    \caption{LDPC $(121,70)$}
    \label{fig_121_70}
\end{subfigure}
\hfill
\begin{subfigure}[b]{.30\textwidth}
    \centering
    \includegraphics[width=\textwidth]{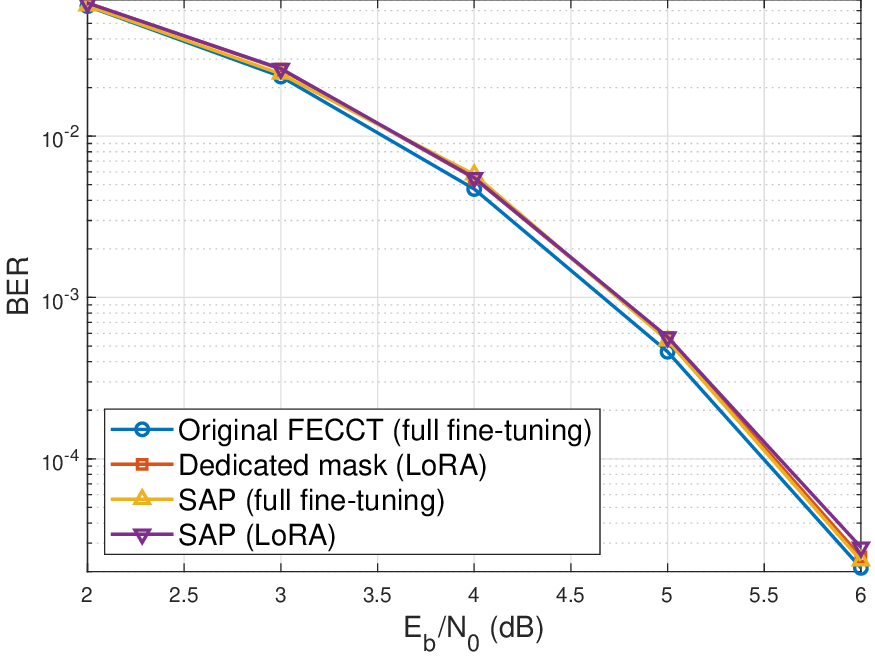}
    \caption{Polar $(128,64)$}
    \label{fig_64_32}
\end{subfigure}
\caption{Decoded BER comparison of the pretrained FECCT baseline, the dedicated pruned models, and the SAP models.    
\label{fig:ber_comparison}
}
\end{center}
\vskip -0.3in
\end{figure*} 

\subsection{Decoding Performance} \label{sec:decoding-performance}

We evaluate SAP on various BCH, LDPC, and polar codes by comparing the following decoder configurations:
(i) an unpruned retraining baseline, in which the pretrained FECCT is fully fine-tuned for the same number of epochs as the recovery stage;
(ii) a dedicated pruned model followed by LoRA recovery; 
(iii) a SAP model followed by full fine-tuning; and
(iv) a SAP model followed by LoRA-based recovery. 
Unless otherwise stated, all pruned models use the same pruning ratio and recovery budget. 
Accordingly, our evaluation considers two factors: pruning strategy (dedicated vs.\ SAP-selected) and recovery method (full fine-tuning vs.\ LoRA).

Table~\ref{tab:decoding_performance} compares dedicated and SAP masks across BCH, LDPC, and polar codes at \(E_b/N_0\in\{4,5,6\}\,\si{\decibel}\).
We report \(-\ln(\mathrm{BER})\) and the gap \(\Delta=\text{SAP}-\text{Dedicated}\).
Across code families, SNRs, and both seen and unseen targets, \(\Delta\) remains close to zero, with most entries satisfying \(\Delta\ge -0.15\).
These results show that SAP-selected shared pruning masks achieve decoding performance comparable to code-specific dedicated pruning masks across different block lengths and rates, while avoiding per-target pruning mask derivation.
For completeness, unpruned FECCT full fine-tuning results are reported in Appendix~\ref{app:full_finetune_results}.
Additional results on Turbo and Reed-Muller codes, which belong to code families absent from the library, are provided in Appendix~\ref{app:novel_families}.

Figure~\ref{fig:ber_comparison} compares BER curves for the evaluated decoder variants.
For BCH $(63,45)$, LDPC $(121,70)$, and polar $(128,64)$, the dedicated and SAP curves closely track the unpruned retraining baseline, indicating that structured pruning preserves the decoding performance.
Moreover, the SAP (LoRA) curve closely matches that of SAP (full fine-tuning), suggesting that code-specific recovery can be effectively achieved via parameter-efficient adaptation on top of the pruned model.
Additional details on the LoRA hyperparameters are provided in Appendix~\ref{app:ablation_lora}.
Taken together, these results show that spectrum-guided pruning mask reuse enables structured pruning for FECCT decoders without material accuracy loss: SAP matches dedicated per-code pruning after recovery, and the recovered pruned models closely track the unpruned retraining baseline under the same training schedule.
We also observe the same qualitative trends under other pruning ratios (see Appendix~\ref{app:pruning_ratio}).
We further report complementary evaluations on frame error rate (FER), also referred to as block error rate (BLER) (see Appendix~\ref{app:fer_ablation}).

\begin{figure*}[!t]
    \centering
    \begin{minipage}[t]{0.65\textwidth} 
        \centering
        \begin{subfigure}[t]{0.49\linewidth} 
            \centering
            \includegraphics[width=\linewidth]{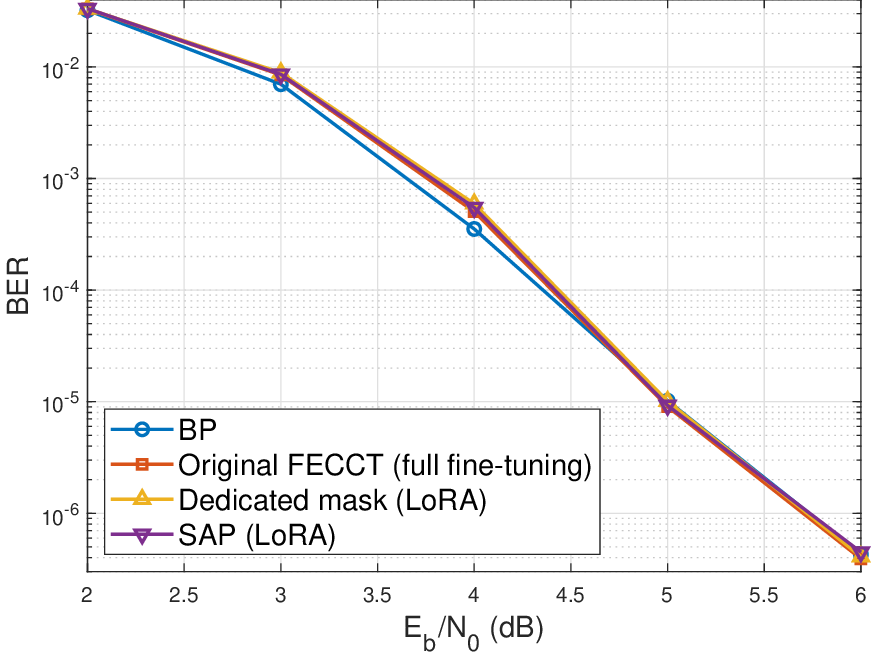}
            \caption{LDPC $(256,224)$}
            \label{fig:long_ldpc_256_224}
        \end{subfigure}
        \hfill 
        \begin{subfigure}[t]{0.49\linewidth}
            \centering
            \includegraphics[width=\linewidth]{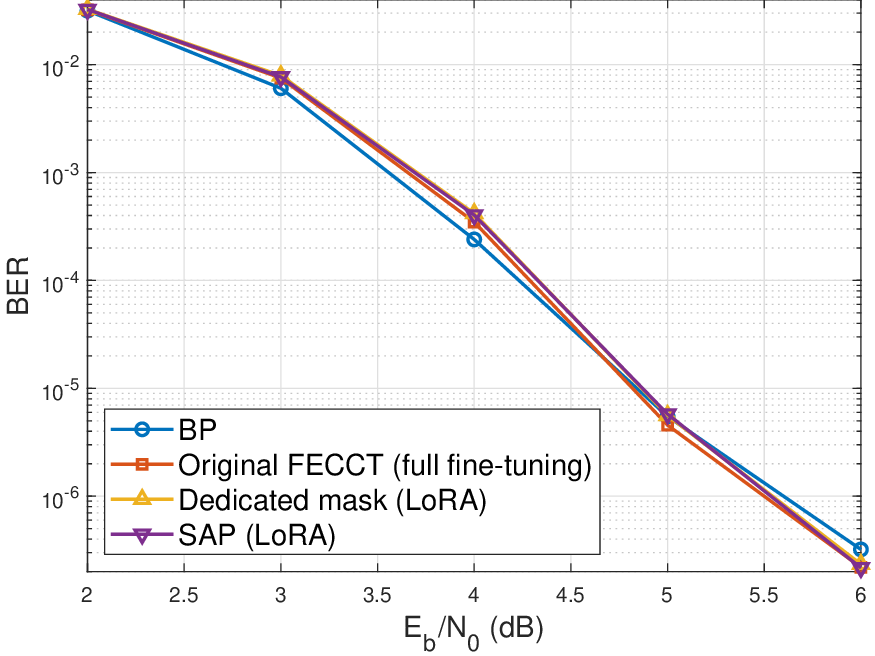}
            \caption{LDPC $(320,296)$}
            \label{fig:long_ldpc_320_296}
        \end{subfigure}
        
        \caption{Decoded BER on 5G NR LDPC codes.
        We compare BP with $50$ iterations, dedicated per-code pruning with recovery, and SAP mask reuse with recovery.}
        \label{fig:long_code_results}
    \end{minipage}
    \hfill 
    \begin{minipage}[t]{0.32\textwidth}
        \centering
        \includegraphics[width=\linewidth]{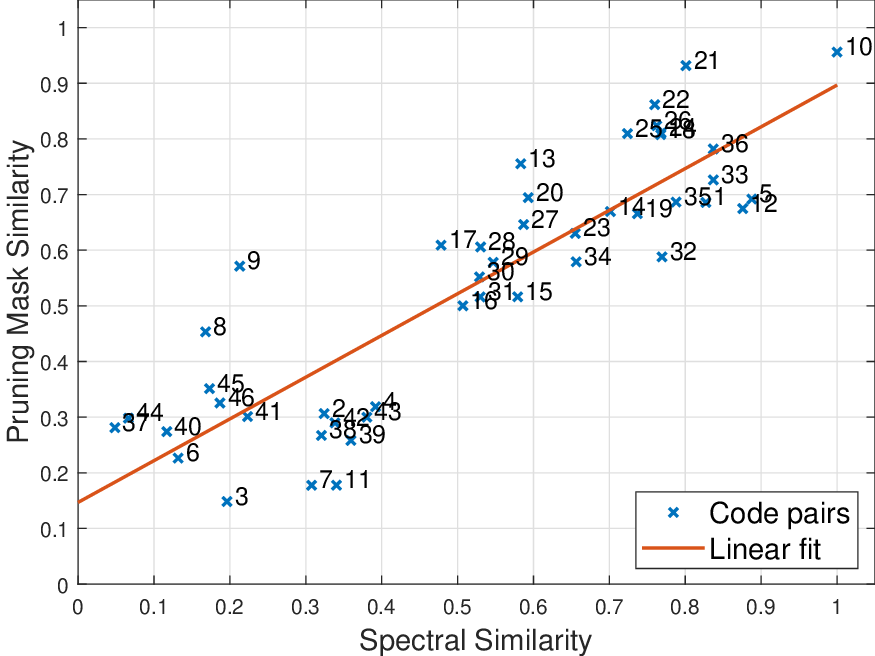}
        \vspace{4pt}
        \caption{Spectral similarity vs.\ pruning mask similarity. Pearson correlation: $\rho=0.87$.}
        \label{fig:spectrum_vs_mask}
    \end{minipage}
\vskip -0.13in
\end{figure*}

\begin{figure*}[t!]
\begin{center}
\begin{subfigure}[b]{.33\textwidth}
    \centering
    \includegraphics[width=\textwidth]{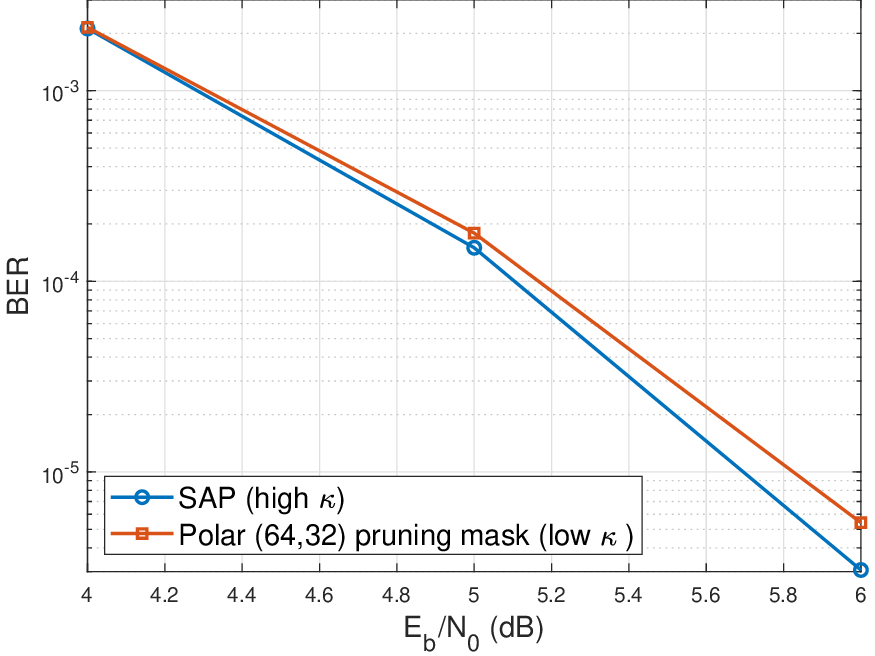}
    \caption{LDPC $(49,24) $}    
\end{subfigure}
\hfil
\begin{subfigure}[b]{.33\textwidth}
    \centering
    \includegraphics[width=\textwidth]{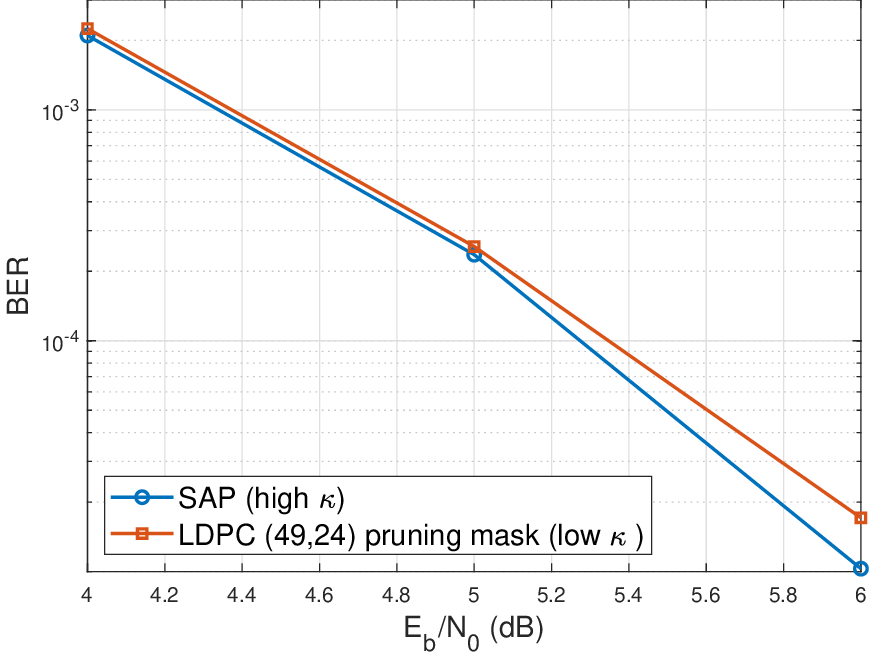}
    \caption{Polar $(64,32)$}
\end{subfigure}

\caption{Pruning mask reuse under low spectral similarity ($\kappa = 0.2978$). 
Reusing a pruning mask derived from a spectrally dissimilar code (LDPC $(49,24)$ and polar $(64,32)$) degrades BER performance relative to SAP's library-selected mask.
\label{fig:cross_transfer}}
\end{center}
\vskip -0.3in
\end{figure*}

\subsection{5G NR LDPC Codes}
\label{app:long_code}

To evaluate SAP on practical standardized codes, we conduct experiments on 5G NR LDPC codes.
For each target code, SAP performs spectral retrieval over the library and selects LDPC $(96,64)$ as the nearest neighbor, with similarities $\kappa^\star=0.8816$ for LDPC $(256,224)$ and $\kappa^\star=0.8817$ for LDPC $(320,296)$.
We reuse the structured pruning mask of LDPC $(96,64)$ and apply the same recovery procedure as in the main experiments.

Figure~\ref{fig:long_code_results} compares the standard belief propagation (BP) decoder, the unpruned retraining baseline of FECCT, and pruned FECCT decoders after recovery using dedicated and SAP-selected masks. 
Across both code settings, the SAP curves closely track those of the dedicated pruning over the tested $E_b/N_0$ range, indicating that the proposed SAP is effective for 5G NR LDPC codes.
Moreover, compared to BP, the unpruned retraining baseline and the recovered pruned decoders (dedicated and SAP) achieve comparable decoded BER. 
Overall, these results show that SAP provides effective pruning mask reuse for 5G NR LDPC codes with longer block lengths without sacrificing decoding performance.
Additional results on longer 5G NR LDPC codes are provided in Appendix~\ref{app:long_nr_ldpc}.
\section{Analysis}\label{sec:analysis}

\subsection{Spectral Similarity Tracks Mask Overlap}
\label{sec:analysis-spectrum-mask}

Figure~\ref{fig:spectrum_vs_mask} illustrates how spectral similarity~\eqref{eq:spec_sim} between bipartite graphs predicts pruning mask similarity over 46 code pairs spanning BCH, LDPC, 5G NR LDPC, and polar codes with diverse block lengths and rates.
The pruning mask similarity is measured using the Jaccard similarity between structured pruning masks.
For readability, we annotate each point with an index, and the corresponding list of code pairs is provided in Appendix~\ref{app:code_pairs}.

The results reveal a strong positive relationship: code pairs with higher spectral similarity tend to yield more similar pruning masks.
A linear fit captures this trend well, with a Pearson correlation coefficient of $\rho=0.87$, indicating a tight alignment between spectral similarity and pruning mask overlap.
This empirical result supports the design of SAP, showing that the bipartite graph spectrum serves as a compact and scalable criterion for identifying which previously computed pruning mask is likely to be compatible with a new code.

We additionally compare spectral similarity with alternative structural similarity measures and simple baselines (see Appendix~\ref{app:alt_similarity}).
The proposed spectral similarity provides the most consistent alignment with pruning mask overlap among the evaluated measures.
In particular, a binary family-label predictor and a code-rate predictor reach only $\rho=0.68$ and $\rho=-0.27$, respectively, so the relation is not explained by code-family membership or code rate alone.

\subsection{Pruning Mask Reuse Fails Under Low Spectral Similarity}
\label{sec:analysis-cross-transfer}

This section examines the failure of naive pruning mask reuse when code structures are highly mismatched.
We consider two codes from different families, LDPC $(49,24)$ and polar $(64,32)$ codes, which exhibit low spectral similarity ($\kappa=0.2978$).
To isolate the effect of pruning mask selection, we conduct a cross-transfer experiment in which each target code is pruned using the other code's structured pruning mask under the same FLOPs budget, followed by the same recovery procedure used in SAP.

As shown in Figure~\ref{fig:cross_transfer}, reusing a pruning mask from a spectrally dissimilar code consistently degrades decoding performance compared to the SAP-selected mask, even after recovery.
Since both the pruning ratio and the recovery procedure are held fixed, the observed performance loss is directly attributable to the mismatch between the pruning mask and the target code's bipartite graph structure.
These results indicate that structured pruning masks in FECCT are not universally reusable across codes with low spectral similarity. 
Selecting a graph-aligned pruning mask is therefore crucial for preserving decoding performance, and SAP's spectrum-based criterion provides a principled and effective mechanism for mask reuse across structurally diverse codes. 

\section{Conclusion}

We presented SAP, a structure-aware pruning framework for universal transformer-based ECC decoders.
SAP reduces per-code pruning overhead by enabling pruning mask reuse across spectrally similar codes.
Extensive evaluations on BCH, LDPC, and polar codes show that spectral similarity strongly correlates with pruning mask overlap, validating SAP's threshold-based reuse decision and on-demand library expansion. 
By combining pruning-mask sharing with parameter-efficient LoRA-based recovery, SAP achieves decoding performance comparable to dedicated per-code pruning while substantially reducing computational cost and model memory footprint. 
Future work includes integrating SAP with complementary efficiency techniques such as quantization and architecture-level sparsification. 

\begin{ack}
This work was supported by the Institute of Information \& Communications Technology Planning \& Evaluation (IITP) grants funded by the Korean government (MSIT) under Grant No.~RS-2024-00398449 (Network Research Center: Advanced Channel Coding and Channel Estimation Technologies for Wireless Communication Evolution) and Grant No.~RS-2026-25523842, and by the BK21 FOUR program. The authors declare no competing interests.
\end{ack}

\bibliographystyle{plainnat}
\bibliography{reference}

\newpage
\appendix

\section{SAP Mask Selection Algorithm}
\label{app:MS_algorithm}

\begin{algorithm}[h]
  \caption{Spectral-Aligned Pruning Mask Selection}
  \label{alg:sap}
  \begin{algorithmic}[1]
    \State \textbf{Input:} Target PCM $H_{\text{new}}$, pruning mask library $\mathcal{D}=\{(\phi(H_i),M(H_i))\}_{i=1}^{N}$, similarity threshold $\tau$
    \State Compute spectral signature $\phi(H_{\text{new}})\in\mathbb{R}^2$
    \For{$i=1$ to $N$}
        \State $d_i \leftarrow d(\phi(H_{\text{new}}), \phi(H_i))$
    \EndFor
    \State $i^\star \leftarrow \arg\min_i d_i$, $d^\star \leftarrow \min_i d_i$, $\kappa^\star \leftarrow \kappa(H_{\text{new}},H_{i^\star})$
    \If{$\kappa^\star \ge \tau$}
      \State $\widehat{M}(H_{\text{new}}) \leftarrow M(H_{i^\star})$
    \Else
      \State Compute a new code-specific mask $M(H_{\text{new}})$
      \State $\widehat{M}(H_{\text{new}}) \leftarrow M(H_{\text{new}})$, 
      $\mathcal{D} \leftarrow \mathcal{D} \cup \{(\phi(H_{\text{new}}), M(H_{\text{new}}))\}$
    \EndIf
    \State \textbf{Output:} Selected mask $\widehat{M}(H_{\text{new}})$; updated library $\mathcal{D}$
  \end{algorithmic}
\end{algorithm}

Algorithm~\ref{alg:sap} summarizes the threshold-based mask selection procedure described in Section~\ref{sec:library}.
SAP reuses the nearest library mask when the spectral similarity exceeds $\tau$; otherwise, it derives a new code-specific mask and inserts the resulting signature-mask pair into the library.

\section{Bipartite Graph Spectrum}
\label{app:spectrum_ecc}

This appendix provides additional background on why SAP uses the adjacency spectrum of the bipartite graph induced by the parity-check matrix for pruning-mask retrieval.
As discussed in Section~\ref{sec:spectrum}, a parity-check matrix \(H\) defines a bipartite graph consisting of variable nodes and check nodes.
The adjacency matrix of this bipartite graph is given by
\[
A(H)=
\begin{bmatrix}
0 & H^\top \\
H & 0
\end{bmatrix}.
\]
In FECCT, the code-dependent attention connectivity is constructed from the distance structure of this bipartite graph.
Therefore, \(A(H)\) represents the code-dependent graph structure used by FECCT.

Different codes can have different block lengths, rates, and parity-check structures, and therefore the size and sparsity pattern of \(A(H)\) also vary across codes.
SAP therefore does not directly compare \(A(H)\) in matrix form.
Instead, it constructs a compact spectral signature from the adjacency spectrum of \(A(H)\).
In particular, SAP uses the two algebraically largest eigenvalues, \(\lambda_1\) and \(\lambda_2\).
This choice is motivated by three graph-spectral perspectives: degree scale, expansion-related connectivity, and graph-based minimum-distance analysis.
Section 3.1 and this appendix provide theoretical motivation for using the two leading adjacency eigenvalues as a compact descriptor of the parity-check bipartite graph, and Sections 4 and 5.1 empirically support this choice. 
This motivation is not limited to regular codes: for irregular codes, the degree-scale analysis in Appendix~\ref{app:DS} and the minimum-distance bounds in Appendix~\ref{app:MDB} show that $\lambda_1$ and $\lambda_2$ remain meaningful descriptors.

\subsection{Degree Scale}\label{app:DS}

The largest adjacency eigenvalue reflects the degree scale of the graph.
For a connected graph with largest eigenvalue \(\lambda_1\), \citep[Proposition~3.1.2]{Brouwer2011spectra} states that \(\lambda_1=d\) for a \(d\)-regular graph, while for a non-regular graph,
\[
\bar{d}<\lambda_1<d_{\max},
\]
where \(\bar{d}\) and \(d_{\max}\) denote the average and maximum degrees, respectively.
Thus, \(\lambda_1\) captures degree-scale information.
This is relevant for parity-check-induced bipartite graphs because node degrees determine how many parity constraints are aggregated at each variable or check node, thereby affecting both BP message passing and graph-conditioned attention connectivity.

\subsection{Expansion Ratio}

The spectral gap characterizes expansion in regular graphs.
Specifically, for a \(d\)-regular graph with eigenvalues \(\lambda_1=d \ge \lambda_2 \ge \cdots \ge \lambda_n\), the spectral gap
\[
\delta = \lambda_1-\lambda_2
\]
bounds the edge expansion ratio \(h(G)\) through Cheeger-type inequalities~\citep[Theorem~2.4]{Hoory2006expander}:
\[
\frac{\delta}{2}
\le
h(G)
\le
\sqrt{2d\delta}.
\]
Thus, a larger spectral gap indicates stronger expansion, meaning that neighborhoods grow more effectively across the graph.
This is relevant to ECC decoding because reliability and constraint information propagate through the bipartite graph during BP message passing or graph-conditioned attention~\citep{Sipser1996Expander}.

\subsection{Minimum-distance Bounds}\label{app:MDB}

The two leading eigenvalues appear in graph-based minimum-distance bounds.
While degree scale and expansion-related connectivity provide graph-theoretic motivation, graph-based distance analysis further shows that the two leading eigenvalues are connected to coding-theoretic quantities.
In particular, minimum-distance lower bounds can be derived from the bipartite graph, using the subgraph associated with a minimum-weight codeword and the leading eigenvalues of the corresponding graph-derived matrices~\citep{Tanner2001minimum}.

For regular codes, this relationship is particularly explicit.
Let \(H\in\{0,1\}^{n-k\times n}\) be a regular parity-check matrix with fixed column weight \(\gamma\) and fixed row weight \(\rho\).
Let \(\mu_1\) and \(\mu_2\) denote the largest and second-largest nonzero eigenvalues of \(H^\top H\) or equivalently \(HH^\top\).
The bit-oriented bound in \citep{Tanner2001minimum} gives the following lower bound on the minimum distance \(d_{\min}\):
\[
d_{\min}
\ge
\frac{n(2\gamma-\mu_2)}{\mu_1-\mu_2}.
\]
The parity-oriented bound is given by
\[
d_{\min}
\ge
\frac{2n(2\gamma+\rho-2-\mu_2)}
{\rho(\mu_1-\mu_2)}.
\]
Here, $\mu_i$ is the eigenvalue of $H^\top H$, which satisfies
\[
\mu_i=\lambda_i^2.
\]

For irregular codes, corresponding minimum-distance lower bounds have also been derived by extending Tanner's graph-based analysis~\citep{Shin2005Generalization}.

SAP does not directly compute or optimize these minimum-distance bounds.
Rather, these bounds provide coding-theoretic motivation for using the two leading adjacency eigenvalues as a compact graph descriptor.
Together with the degree-scale and expansion-ratio interpretations, these minimum-distance bounds further support the choice of the two-dimensional spectral signature \(\phi(H)=[\lambda_1,\lambda_2]\) in SAP.

\section{Code-Pair Indices for Figure~\ref{fig:spectrum_vs_mask}}
\label{app:code_pairs}

For readability, Figure~\ref{fig:spectrum_vs_mask} annotates each scatter point with a numeric index rather than a full code-pair label.
Table~\ref{tab:pair_index} lists the corresponding 46 code pairs, which span BCH, conventional LDPC, 5G NR LDPC, and polar codes with diverse block lengths and rates.
Most entries compare two different codes, while indices 10-12 are controlled pairs constructed from the same nominal code.
These controlled pairs are included to examine whether pruning-mask similarity is governed only by the code parameters $(n,k)$, or also by the specific parity-check matrix and the resulting bipartite graph structure.

\begin{table}[h]
\caption{Code-pair indices used in Figure~\ref{fig:spectrum_vs_mask}.
Indices marked with $\dagger$ are controlled pairs built from the same nominal code with different parity-check matrices (for index 10, different edge connections).}
\label{tab:pair_index}
\vskip 0.15in
\centering
\footnotesize
\setlength{\tabcolsep}{4pt}
\begin{tabular}{cl@{\hspace{14pt}}cl}
\toprule
\textbf{Index} & \textbf{Code pair} & \textbf{Index} & \textbf{Code pair} \\
\midrule
1  & LDPC $(49,24)$ vs. LDPC $(121,80)$   & 24 & LDPC $(96,72)$ vs. LDPC $(96,64)$ \\
2  & BCH $(63,36)$ vs. LDPC $(49,24)$     & 25 & LDPC $(96,76)$ vs. LDPC $(96,64)$ \\
3  & Polar $(64,32)$ vs. LDPC $(49,24)$   & 26 & LDPC $(128,96)$ vs. LDPC $(96,64)$ \\
4  & BCH $(63,36)$ vs. LDPC $(121,80)$    & 27 & LDPC $(256,224)$ vs. LDPC $(96,64)$ \\
5  & BCH $(63,45)$ vs. BCH $(63,51)$      & 28 & LDPC $(320,296)$ vs. LDPC $(96,64)$ \\
6  & BCH $(63,36)$ vs. Polar $(128,64)$   & 29 & Polar $(32,11)$ vs. BCH $(31,16)$ \\
7  & Polar $(64,48)$ vs. LDPC $(121,80)$  & 30 & Polar $(32,16)$ vs. BCH $(31,16)$ \\
8  & Polar $(64,48)$ vs. Polar $(128,64)$ & 31 & Polar $(32,22)$ vs. BCH $(31,16)$ \\
9  & Polar $(64,48)$ vs. Polar $(128,86)$ & 32 & Polar $(64,32)$ vs. Polar $(64,48)$ \\
10$^\dagger$ & LDPC $(128,64)$ vs. LDPC $(128,64)$ & 33 & Polar $(64,43)$ vs. Polar $(64,48)$ \\
11$^\dagger$ & LDPC $(49,24)$ vs. LDPC $(49,24)$   & 34 & Polar $(128,43)$ vs. Polar $(128,86)$ \\
12$^\dagger$ & BCH $(63,45)$ vs. BCH $(63,45)$     & 35 & Polar $(128,64)$ vs. Polar $(128,86)$ \\
13 & BCH $(31,11)$ vs. BCH $(31,16)$      & 36 & Polar $(128,96)$ vs. Polar $(128,86)$ \\
14 & BCH $(31,21)$ vs. BCH $(31,16)$      & 37 & LDPC $(32,16)$ vs. Polar $(128,64)$ \\
15 & BCH $(31,26)$ vs. BCH $(31,16)$      & 38 & LDPC $(96,48)$ vs. Polar $(32,16)$ \\
16 & BCH $(63,30)$ vs. LDPC $(96,64)$     & 39 & LDPC $(128,64)$ vs. Polar $(64,32)$ \\
17 & BCH $(63,36)$ vs. BCH $(63,51)$      & 40 & LDPC $(64,48)$ vs. Polar $(128,96)$ \\
18 & LDPC $(32,16)$ vs. BCH $(31,16)$     & 41 & LDPC $(128,96)$ vs. Polar $(32,24)$ \\
19 & LDPC $(49,24)$ vs. LDPC $(121,60)$   & 42 & LDPC $(96,72)$ vs. Polar $(64,48)$ \\
20 & LDPC $(96,48)$ vs. LDPC $(96,64)$    & 43 & LDPC $(96,64)$ vs. Polar $(64,43)$ \\
21 & LDPC $(121,70)$ vs. LDPC $(121,60)$  & 44 & BCH $(31,21)$ vs. Polar $(128,86)$ \\
22 & LDPC $(121,80)$ vs. LDPC $(121,60)$  & 45 & LDPC $(320,296)$ vs. LDPC $(48,44)$ \\
23 & LDPC $(64,48)$ vs. LDPC $(96,64)$    & 46 & BCH $(127,113)$ vs. LDPC $(256,224)$ \\
\bottomrule
\end{tabular}
\end{table}

\section{Implementation Details}
\label{app:implementation_details}

Table~\ref{tab:implementation_details} summarizes the implementation settings used in all experiments.
SAP is applied to a FECCT backbone pretrained on the code set in Appendix~\ref{app:pretrain_codes}.
For each target code, SAP selects or derives a pruning mask, physically removes the pruned attention heads and FFN channels, and then recovers decoding performance with LoRA.

\begin{table}[h]
\caption{Implementation details of the FECCT backbone, pruning, the SAP library, and recovery.}
\label{tab:implementation_details}
\vskip 0.15in
\centering
\small
\setlength{\tabcolsep}{6pt}
\begin{tabular}{llp{0.50\linewidth}}
\toprule
\textbf{Component} & \textbf{Item} & \textbf{Setting} \\
\midrule
\multirow{4}{*}{Backbone}
 & Decoding layers         & $N=6$ \\
 & Embedding dimension     & $d=128$ \\
 & Self-attention heads    & 8 \\
 & FFN hidden dimension    & $d_{\mathrm{FFN}}=512$ \\
\midrule
\multirow{5}{*}{Pretraining}
 & Codes                   & 12 codes in Appendix~\ref{app:pretrain_codes} \\
 & Training data           & All-zero codewords \\
 & Epochs / batch size     & 4{,}000 / 256 \\
 & $E_b/N_0$               & Uniformly sampled from \SIrange{2}{7}{\decibel} \\
 & Optimizer / schedule    & Adam / cosine decay from $1\times10^{-4}$ to $1\times10^{-6}$ \\
\midrule
\multirow{4}{*}{Pruning}
 & Importance estimator    & Fisher-based~\citep{Kwon2022fast} \\
 & Calibration set         & 128 samples generated at $E_b/N_0=\SI{4}{\decibel}$ \\
 & Pruned units            & Attention heads and FFN channels \\
 & Pruning ratio           & \SI{40}{\percent} (FLOPs-based), with MAC-constrained structured search \\
\midrule
\multirow{3}{*}{SAP library}
 & Spectral signature      & $\phi(H)=[\lambda_1,\lambda_2]$, with $\beta=0.1$ \\
 & Reuse threshold         & $\tau=0.5$ \\
 & Initial library         & Six reference codes \\
\midrule
\multirow{5}{*}{Recovery}
 & LoRA                    & Rank $r=8$, scaling factor $\alpha=16$ \\
 & Target modules          & $W_Q$, $W_K$, $W_V$, $W_O$ in all attention layers \\
 & Epochs                  & 100 \\
 & Loss                    & BCE $+\ \gamma\cdot$ KD, with $\gamma=1$ \\
 & Knowledge distillation  & Teacher: pretrained unpruned FECCT; average KL divergence between bitwise posteriors; no temperature scaling ($T=1$) \\
\midrule
Hardware & GPU & NVIDIA RTX A6000 \\
\bottomrule
\end{tabular}
\end{table}

\section{Pretraining Code Set}\label{app:pretrain_codes}

\begin{table*}[!h]
\caption{Pretraining code set for the FECCT backbone.}
\label{tab:pretrain_codes}
\begin{center}
\begin{small}
\begin{tabular}{l l}
\toprule
Family & Codes used for pretraining \\
\midrule
BCH   & $(31,16)$, $(63,36)$, $(63,45)$, $(63,51)$ \\
Polar & $(64,32)$, $(64,48)$, $(128,64)$, $(128,86)$ \\
LDPC  & $(49,24)$, $(121,60)$, $(121,70)$, $(121,80)$ \\
\bottomrule
\end{tabular}
\end{small}
\end{center}
\vskip -0.1in
\end{table*}

Table~\ref{tab:pretrain_codes} lists the 12 codes used to pretrain the FECCT backbone, covering three representative ECC families (BCH, LDPC, and polar).
Following common practice in transformer-based ECC decoder (ECCT/FECCT) evaluations, we select representative codes spanning a range of rates and block lengths within each family~\citep{Choukroun2022error,Choukroun2024foundation}.
This set is used throughout our experiments unless stated otherwise.

\section{Wall-Clock Cost of Fisher-Based Mask Derivation vs. Reuse}
\label{app:mask_cost}

\begin{table}[h]
\centering
\caption{Wall-clock cost of spectral-signature-based mask reuse and Fisher-based new-mask derivation.}
\label{tab:mask_cost}
\begin{tabular}{lcc}
\toprule
Codes & Mask reuse & New mask derivation \\
\midrule
BCH $(31,16)$    & $0.00022$ s & $47.83$ s \\
BCH $(63,45)$    & $0.00039$ s & $55.51$ s \\
LDPC $(96,64)$   & $0.00087$ s & $120.88$ s \\
LDPC $(121,70)$  & $0.00136$ s & $160.71$ s \\
Polar $(64,32)$  & $0.00046$ s & $136.27$ s \\
Polar $(128,64)$ & $0.00022$ s & $181.47$ s \\
\bottomrule
\end{tabular}
\end{table}

The SAP pipeline contains two computationally different paths.
Deriving a new mask requires Fisher-based importance estimation, which involves forward and backward passes on calibration data followed by a MAC-constrained structured search.
In contrast, mask reuse only requires computing the target code's bipartite-graph adjacency spectrum, retrieving the nearest spectral signature from the library, and applying the threshold decision.
This distinction is directly aligned with the core motivation of SAP: amortizing the cost of structured pruning-mask derivation across codes.

To quantify this difference, we separately measure the wall-clock time of Fisher-based new-mask derivation and spectral-signature-based mask reuse on several representative codes.
As shown in Table~\ref{tab:mask_cost}, mask reuse is several orders of magnitude faster than deriving a new mask.
These results support the practical advantage of SAP: it substantially reduces the overhead of pruning-mask search not only algorithmically, but also in wall-clock time.

\section{Memory Footprint and Inference Latency after Structured Pruning}
\label{app:memory_footprint}

\subsection{Memory Footprint}

\begin{table}[h]
\centering
\caption{Comparison of memory footprint between the full FECCT and its structured-pruned variant across multiple codes.
All pruned models use a \SI{40}{\percent} pruning ratio.
Red.(\%) \(=100\times(\text{Full}-\text{Pruned})/\text{Full}\).}
\label{tab:memory_footprint}
\vskip 0.15in
{\begin{tabular}{lccc}
\toprule
\multirow{2}{*}{Code} & \multicolumn{3}{c}{Memory Footprint (MB)} \\
\cmidrule(lr){2-4}
& Full & Pruned & Red. (\%) \\
\midrule
BCH $(31,16)$    & 4.67 & 2.79 & 40.26 \\
BCH $(63,45)$    & 4.67 & 2.74 & 41.33 \\
LDPC $(96,64)$   & 4.67 & 2.67 & 42.83 \\
LDPC $(121,70)$  & 4.67 & 2.62 & 43.90 \\
Polar $(64,32)$  & 4.67 & 2.69 & 42.40 \\
Polar $(128,64)$ & 4.67 & 2.55 & 45.40 \\
\bottomrule
\end{tabular}}
\vskip -0.10in
\end{table}

Table~\ref{tab:memory_footprint} reports the memory footprint of the full FECCT and the structurally pruned model across representative codes.
All pruned models use the same \SI{40}{\percent} FLOPs-based pruning ratio as in the main experiments.
After pruning, the selected attention heads and FFN channels are physically removed from the model, reducing the memory footprint of the pruned model.

Across the evaluated codes, structured pruning reduces the memory footprint from \(4.67\) MB to \(2.55\)-\(2.79\) MB, corresponding to a reduction of \(40.26\%\)-\(45.40\%\).
The reduction is slightly code-dependent because the effective contribution of attention and FFN components varies with the code length and the resulting sequence size.
These results complement the FLOPs and parameter-count analysis in Table~\ref{tab:efficiency}, showing that SAP provides not only computational savings but also a direct reduction in the memory footprint of the pruned model.

\subsection{Inference Latency and Throughput}
\label{app:latency}

FLOPs and memory footprint do not fully determine practical inference efficiency, so we also measure the wall-clock latency and throughput of the full FECCT and the structurally pruned model on a GPU.
After 50 warm-up iterations, we measure 200 forward passes using CUDA events with explicit synchronization, with a batch of 128 codewords.
The per-codeword latency is the measured batch latency divided by 128, and the throughput is the number of decoded codewords divided by the total wall-clock time.
All inputs are prepared on the GPU before timing, and the same protocol is applied to both models.

Table~\ref{tab:latency_throughput} shows that structured pruning reduces the latency by \(12.7\%\)-\(22.7\%\) and increases the throughput by \(14.8\%\)-\(29.6\%\).
These gains are smaller than the \SI{40}{\percent} FLOPs reduction because operations that are not removed by pruning also contribute to the runtime.
The gain is largest for the longest code, LDPC $(512,474)$.

\begin{table}[h]
\centering
\caption{End-to-end inference latency and throughput of the full FECCT and its structurally pruned model (\SI{40}{\percent} pruning ratio).
Latency is reported per codeword, measured with a batch of 128 codewords.
Red.\ and Impr.\ denote the latency reduction and the throughput improvement of the pruned model, respectively.}
\label{tab:latency_throughput}
\vskip 0.15in
\centering
\small
\setlength{\tabcolsep}{5pt}
{\begin{tabular}{lcccccc}
\toprule
\multirow{2}{*}{Code} & \multicolumn{3}{c}{Latency (\si{\micro\second})} & \multicolumn{3}{c}{Throughput (codewords/s)} \\
\cmidrule(lr){2-4}\cmidrule(lr){5-7}
& Full & Pruned & Red. (\%) & Full & Pruned & Impr. (\%) \\
\midrule
LDPC $(512,474)$ & 790.00 & 610.54 & 22.72 & 1264.3  & 1638.9  & 29.63 \\
BCH $(63,45)$    & 69.14  & 59.45  & 14.02 & 14622.5 & 17021.1 & 16.40 \\
LDPC $(121,70)$  & 216.79 & 189.29 & 12.68 & 4629.2  & 5313.5  & 14.78 \\
\bottomrule
\end{tabular}}
\vskip -0.10in
\end{table}

\section{Alternative Similarity Metrics: Why Adjacency Spectrum is Preferred}
\label{app:alt_similarity}

\begin{figure*}[!h]
\begin{center}
\begin{subfigure}[b]{0.32\textwidth}
    \centering
    \includegraphics[width=\textwidth]{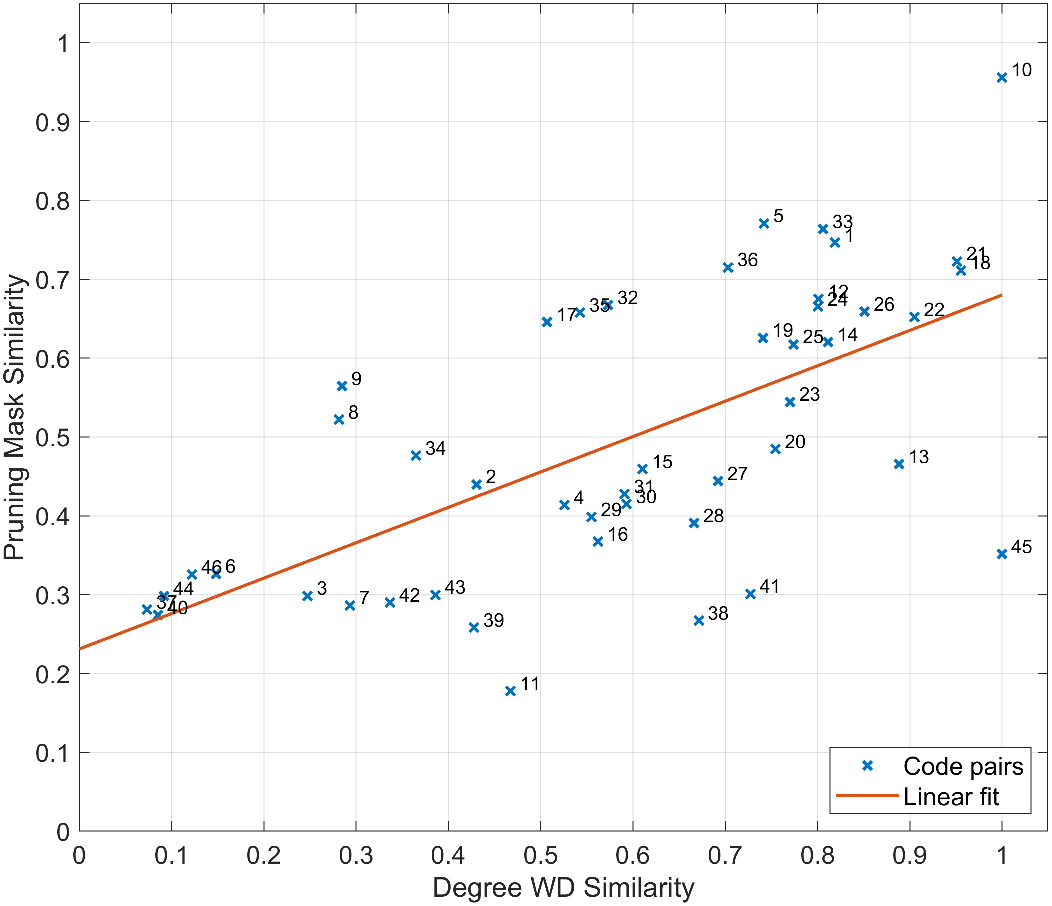}
    \caption{Degree-distribution Wasserstein distance similarity ($\rho=0.64$)}
    \label{fig:wd_sim}
\end{subfigure}
\hfill
\begin{subfigure}[b]{0.32\textwidth}
    \centering
    \includegraphics[width=\textwidth]{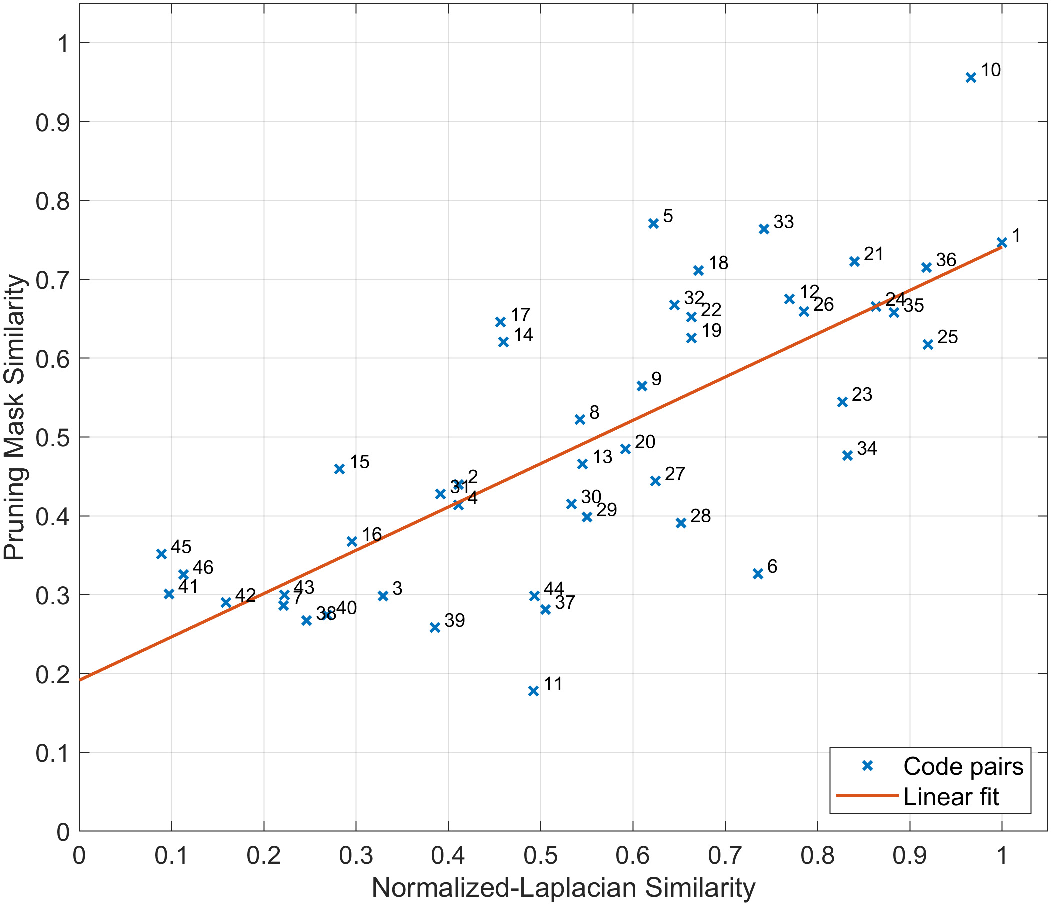}
    \caption{Normalized-Laplacian spectrum similarity ($\rho=0.75$)}
    \label{fig:nor_lap_sim}
\end{subfigure}
\hfill
\begin{subfigure}[b]{0.32\textwidth}
    \centering
    \includegraphics[width=\textwidth]{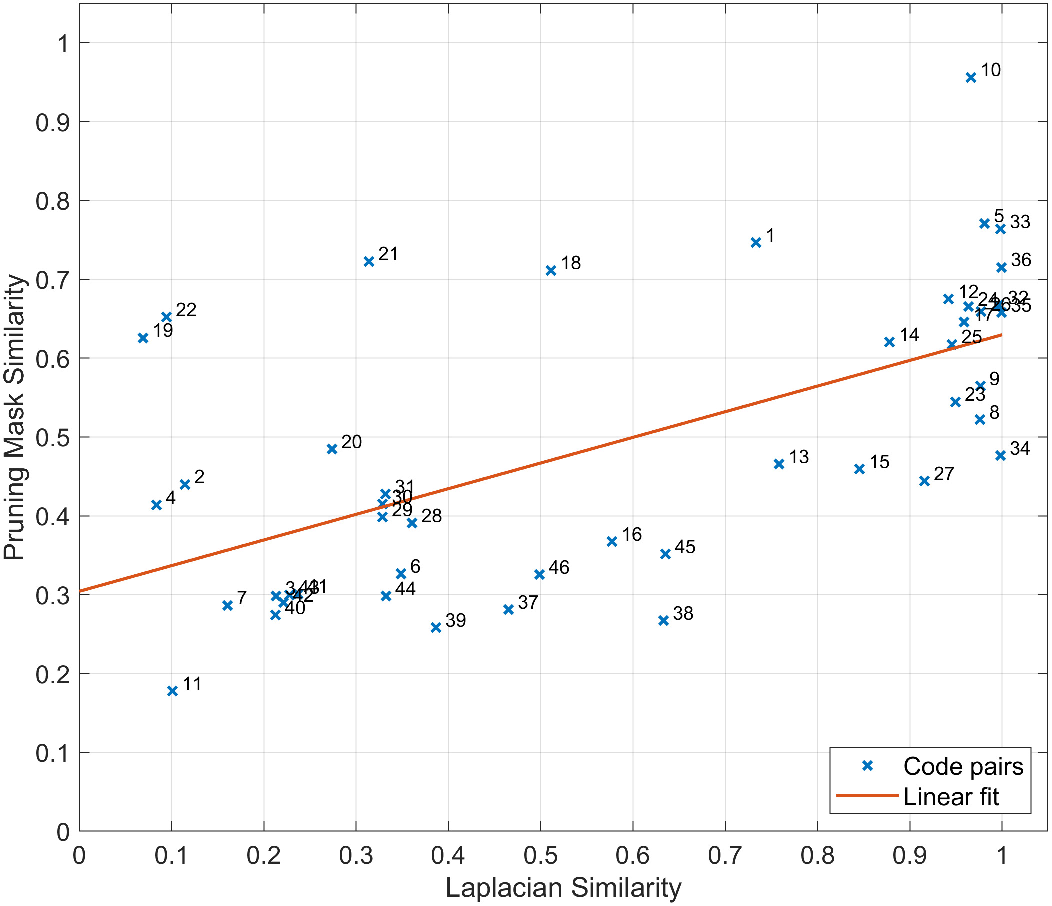}
    \caption{Laplacian spectrum similarity ($\rho=0.61$)}
    \label{fig:lap_sim}
\end{subfigure}
\caption{
Correlation between pruning-mask similarity (Jaccard index) and alternative similarity metrics.
Each point corresponds to a code pair (indexed as in Table~\ref{tab:pair_index}), and the red line is a least-squares linear fit.
}
\label{fig:alt_similarity}
\end{center}
\end{figure*}

We compare the adjacency-spectrum similarity used in SAP with three alternative graph-based similarity metrics: degree-distribution Wasserstein distance similarity, normalized-Laplacian spectrum similarity, and combinatorial Laplacian spectrum similarity.
For each metric, we measure its correlation with pruning-mask similarity, computed as the Jaccard similarity between structured pruning masks.
Figure~\ref{fig:alt_similarity} shows the results for the three alternative metrics.
Degree-distribution Wasserstein similarity achieves a Pearson correlation of \(\rho=0.64\), normalized-Laplacian spectrum similarity achieves \(\rho=0.75\), and Laplacian spectrum similarity achieves \(\rho=0.61\).
All three metrics exhibit positive correlation with pruning-mask similarity, but their correlations are lower than that of the adjacency-spectrum similarity used in the main text, which achieves \(\rho=0.87\).
Thus, in our evaluated setting, adjacency-spectrum similarity provides the most consistent alignment with pruning-mask overlap.

\paragraph{Degree-distribution Wasserstein distance similarity.}
The first alternative uses degree distributions.
Given a parity-check matrix \(H\), we construct the corresponding bipartite graph and compute the edge-perspective degree distributions on the variable-node and check-node sides.
For two codes \(H_A\) and \(H_B\), we measure the discrepancy between the degree distributions on each side using the 1D Wasserstein distance, and average the two distances to obtain \(d_{\mathrm{WD}}\).
We then convert this distance into a similarity score as
\[
\kappa_{\mathrm{WD}}
=
\exp(-\beta d_{\mathrm{WD}}).
\]

\paragraph{Normalized-Laplacian spectrum similarity.}
The second alternative uses the normalized-Laplacian spectrum.
Let \(A\) and \(D\) denote the adjacency matrix and degree matrix of the bipartite graph, respectively.
The normalized Laplacian is defined as
\[
L_{\mathrm{norm}}
=
I-D^{-1/2}AD^{-1/2}.
\]
For each code, we construct a spectral signature \(\phi_{\mathrm{norm}}(H)\) using the smallest non-trivial eigenvalues of \(L_{\mathrm{norm}}\).
The distance and similarity between two codes are defined as
\[
d_{\mathrm{norm}}(\phi_{\mathrm{norm}}(H_A),\phi_{\mathrm{norm}}(H_B))
=
\|\phi_{\mathrm{norm}}(H_A)-\phi_{\mathrm{norm}}(H_B)\|_2,
\]
\[
\kappa_{\mathrm{norm}}
=
\exp(-\beta d_{\mathrm{norm}}).
\]

\paragraph{Laplacian spectrum similarity.}
The third alternative uses the Laplacian spectrum.
The Laplacian is defined as
\[
L
=
D-A.
\]
For each code, we construct a spectral signature \(\phi_{\mathrm{lap}}(H)\) using the smallest non-trivial eigenvalues of \(L\).
The distance and similarity between two codes are defined as
\[
d_{\mathrm{lap}}(\phi_{\mathrm{lap}}(H_A),\phi_{\mathrm{lap}}(H_B))
=
\|\phi_{\mathrm{lap}}(H_A)-\phi_{\mathrm{lap}}(H_B)\|_2,
\]
\[
\kappa_{\mathrm{lap}}
=
\exp(-\beta d_{\mathrm{lap}}).
\]

Overall, degree distributions, normalized-Laplacian spectra, and Laplacian spectra are all reasonable graph-based descriptors for comparing bipartite graph structure.
However, as shown in Figure~\ref{fig:alt_similarity}, all three alternatives show lower correlation with pruning-mask similarity than adjacency spectrum similarity.
Therefore, SAP uses adjacency spectrum similarity as its retrieval metric, considering both the coding-theoretic motivation discussed in the main text and its empirical alignment with pruning mask overlap.

We also compare spectral similarity with simple baselines on the same 46 code pairs.
A binary family-label predictor, which assigns 1 to pairs from the same code family and 0 otherwise, achieves $\rho=0.68$ with pruning-mask similarity, and a code-rate predictor, defined as one minus the absolute difference between the two code rates, achieves $\rho=-0.27$.
Both are lower than the correlation of spectral similarity ($\rho=0.87$).

Cosine similarity between the signatures $[\lambda_1,\lambda_2]$ reaches only $\rho=0.57$.
Since $\lambda_1\ge\lambda_2\ge0$, every signature lies within an angular range of $0^\circ$ to $45^\circ$, and the largest angular difference observed in our evaluation is about $20^\circ$ ($\cos 20^\circ\approx0.94$), so the cosine values concentrate near one.
Moreover, cosine similarity discards the absolute magnitude of $\lambda_1$, which reflects the degree scale of the bipartite graph and is precisely the information that SAP exploits.

\section{Empirical Validation of the Two-Eigenvalue Spectral Signature}
\label{app:k_sensitivity}

\begin{table}[h]
\centering
\caption{Reference-code mapping under adjacency-spectrum retrieval for different spectral signature dimensions $K$. 
Mappings different from the reference code used in the main experiments are highlighted in blue.}
\vskip 0.15in
\label{tab:k_sensitivity}
\resizebox{\textwidth}{!}{
\begin{tabular}{lcccc}
\toprule
\multirow{2}{*}{Target Codes} & \multicolumn{4}{c}{Selected Reference Code} \\
\cmidrule(lr){2-5}
& $K=1$ & $K=2$ & $K=3$ & $K=5$ \\
\midrule
BCH $(31,11)$ 
& BCH $(31,16)$ & BCH $(31,16)$ & BCH $(31,16)$ & BCH $(31,16)$ \\
BCH $(31,21)$ 
& \textcolor{blue}{LDPC $(121,60)$} & BCH $(31,16)$ & BCH $(31,16)$ & BCH $(31,16)$ \\
BCH $(63,36)$ 
& BCH $(63,51)$ & BCH $(63,51)$ & BCH $(63,51)$ & BCH $(63,51)$ \\
BCH $(63,45)$ 
& BCH $(63,51)$ & BCH $(63,51)$ & BCH $(63,51)$ & BCH $(63,51)$ \\
LDPC $(32,16)$ 
& BCH $(31,16)$ & BCH $(31,16)$ & BCH $(31,16)$ & BCH $(31,16)$ \\
LDPC $(64,48)$ 
& \textcolor{blue}{BCH $(31,16)$} & LDPC $(96,64)$ & LDPC $(96,64)$ & LDPC $(96,64)$ \\
LDPC $(96,72)$ 
& LDPC $(96,64)$ & LDPC $(96,64)$ & LDPC $(96,64)$ & LDPC $(96,64)$ \\
LDPC $(49,24)$ 
& LDPC $(121,60)$ & LDPC $(121,60)$ & LDPC $(121,60)$ & LDPC $(121,60)$ \\
LDPC $(121,70)$ 
& LDPC $(121,60)$ & LDPC $(121,60)$ & LDPC $(121,60)$ & LDPC $(121,60)$ \\
Polar $(64,32)$ 
& Polar $(64,48)$ & Polar $(64,48)$ & Polar $(64,48)$ & Polar $(64,48)$ \\
Polar $(64,43)$ 
& Polar $(64,48)$ & Polar $(64,48)$ & Polar $(64,48)$ & Polar $(64,48)$ \\
Polar $(128,64)$ 
& Polar $(128,86)$ & Polar $(128,86)$ & Polar $(128,86)$ & Polar $(128,86)$ \\
Polar $(128,96)$ 
& Polar $(128,86)$ & Polar $(128,86)$ & Polar $(128,86)$ & Polar $(128,86)$ \\
\bottomrule
\end{tabular}
}
\end{table}

Section~\ref{sec:spectrum} and Appendix~\ref{app:spectrum_ecc} motivate SAP's use of the two algebraically largest adjacency eigenvalues from three graph-spectral perspectives: degree scale, expansion-related connectivity, and minimum-distance bounds.
To examine whether this theoretically motivated choice is also empirically stable, we introduce \(K\) as the number of leading adjacency eigenvalues used to form the retrieval signature and evaluate how the selected reference codes change as \(K\) varies.
Specifically, for each \(K \in \{1,2,3,5\}\), we construct a \(K\)-dimensional signature using the \(K\) algebraically largest adjacency eigenvalues and repeat the retrieval stage for the target codes listed in Table~\ref{tab:decoding_performance} under the same library-code setting as in the main experiments.

Table~\ref{tab:k_sensitivity} shows the selected reference code under each value of \(K\).
The results indicate that using only one eigenvalue is insufficient in some cases, as the retrieval mapping changes for several target codes when moving from \(K=1\) to \(K=2\).
In contrast, the retrieval mappings for \(K=2\), \(K=3\), and \(K=5\) are identical for all target codes.

These results support the use of the two-eigenvalue signature \(\phi(H)=[\lambda_1,\lambda_2]\) adopted in SAP.
Increasing \(K\) beyond \(2\) does not change the retrieval mappings in our evaluation, indicating that the theoretically motivated two-eigenvalue signature is empirically stable.
Conversely, using only \(K=1\) is insufficient in some cases, which is consistent with the role of the second eigenvalue in graph-spectral quantities such as expansion-related connectivity and minimum-distance analysis.

\section{Role of Reuse Threshold $\tau$}
\label{app:tau_threshold}

In SAP, $\tau$ is the threshold that controls the mask reuse ratio, and therefore governs the trade-off between decoding accuracy and computational cost.
This role of $\tau$ can be understood through two extreme cases.
If $\tau=0$, any target code can reuse the nearest library mask regardless of how low the similarity score is.
In this case, the cost of deriving a new mask is eliminated, but the risk of inappropriate mask reuse becomes the highest, which may degrade decoding performance.
At the other extreme, if $\tau=1$, reuse is allowed only when the similarity score is exactly one.
This makes SAP the most conservative setting: almost every code receives its own mask.

\subsection{Reuse Rate and Library Growth}
\label{app:tau_library_growth}
To examine how $\tau$ affects mask reuse and library growth, we conduct a stress test in which we sequentially process 32 additional codes starting from the library of six reference codes, applying the update rule in \eqref{eq:reuse_rule} and \eqref{eq:library_update} for each $\tau\in\{0.3,0.5,0.7,0.9\}$.
The reuse rate is the fraction of these codes whose nearest library entry satisfies $\kappa^\star\ge\tau$.
As shown in Table~\ref{tab:tau_library_growth}, the reuse rate decreases and the final library size increases monotonically with $\tau$.
Our default $\tau=0.5$ lies between these extremes: 29 of the 32 codes reuse a mask, and the library grows only from six to nine entries.

Since the final library size increases monotonically with $\tau$, $\tau$ can be set according to the available computational budget.
We sweep $\tau$ over a representative set of candidate codes to obtain the final library size for each value, and the upper bound on the number of masks that can be derived then determines the corresponding $\tau$.

\begin{table}[h]
\caption{Effect of the reuse threshold $\tau$ on mask reuse and library growth.
Thirty-two additional codes are processed sequentially, starting from a library of six reference codes.}
\label{tab:tau_library_growth}
\vskip 0.15in
\centering
\small
\setlength{\tabcolsep}{6pt}
\begin{tabular}{cccc}
\toprule
\textbf{$\tau$} & \textbf{New library entries} & \textbf{Reuse rate (\%)} & \textbf{Final library size} \\
\midrule
0.3 & 0  & 100.0 & 6  \\
0.5 & 3  & 90.6  & 9  \\
0.7 & 16 & 50.0  & 22 \\
0.9 & 30 & 6.3   & 36 \\
\bottomrule
\end{tabular}
\end{table}

Table~\ref{tab:library_clusters} shows the composition of the library at $\tau=0.5$ after all 32 codes are processed.
Entries 1-6 are the reference codes, and entries 7-9 are the three masks newly derived during the stress test.
The library therefore does not grow in proportion to the number of supported codes.
It expands only when a target code is not sufficiently similar to any existing spectral cluster, and it remains compact at the scale of the 38 codes considered here.

\begin{table}[h]
\caption{Codes assigned to each pruning mask library entry at $\tau=0.5$ after processing the 32 additional codes.
Entries 1-6 correspond to the six reference codes of the initial library, and entries 7-9 are masks newly derived during the stress test.}
\label{tab:library_clusters}
\vskip 0.15in
\centering
\small
\setlength{\tabcolsep}{6pt}
\begin{tabular}{cp{0.72\linewidth}}
\toprule
\textbf{Library index} & \textbf{Assigned codes} \\
\midrule
1 & BCH $(31,16)$, $(31,11)$, $(31,21)$, $(31,26)$; LDPC $(32,16)$, $(48,44)$; Polar $(32,11)$, $(32,16)$, $(32,22)$, $(32,24)$ \\
2 & BCH $(63,51)$, $(63,36)$, $(63,45)$ \\
3 & BCH $(63,30)$; LDPC $(96,64)$, $(96,48)$, $(64,48)$, $(96,72)$, $(96,76)$, $(128,96)$, $(256,224)$, $(320,296)$ \\
4 & LDPC $(121,60)$, $(49,24)$, $(121,70)$, $(121,80)$ \\
5 & BCH $(127,120)$; Polar $(64,48)$, $(64,32)$, $(64,43)$ \\
6 & Polar $(128,86)$, $(128,43)$, $(128,64)$, $(128,96)$ \\
7 & BCH $(63,39)$; Polar $(64,22)$ \\
8 & BCH $(127,113)$ \\
9 & LDPC $(128,64)$ \\
\bottomrule
\end{tabular}
\end{table}

\subsection{Decoding Performance under Aggressive Reuse}
\label{app:tau_reliability}

Lowering $\tau$ reduces library growth, but it may allow reuse of a less compatible mask.
To illustrate this trade-off, we select three codes whose nearest library mask is reused at $\tau=0.3$ but rejected at $\tau=0.5$, so that a new dedicated mask is derived.
For each code, we compare the reused mask with the derived dedicated mask after the same LoRA recovery.
As shown in Table~\ref{tab:tau_reliability}, reuse at $\tau=0.3$ lowers $-\ln(\mathrm{BER})$ at 5~dB and 6~dB for all three codes.
Since these codes are borderline cases chosen between the two thresholds, the results show the loss that aggressive reuse can cause rather than its average effect.
They confirm that $\tau=0.5$ avoids such reuse by expanding the library.

\begin{table}[h]
\caption{Decoding performance of a reused library mask ($\tau=0.3$) and a newly derived dedicated mask ($\tau=0.5$) for three borderline codes.
We report $-\ln(\mathrm{BER})$ at $E_b/N_0\in\{4,5,6\}$~\si[]{\decibel} (larger is better).
Both masks are followed by the same LoRA recovery.}
\label{tab:tau_reliability}
\vskip 0.15in
\centering
\small
\setlength{\tabcolsep}{6pt}
\begin{tabular}{llccc}
\toprule
\textbf{Code} & \textbf{Mask} & \textbf{4 dB} & \textbf{5 dB} & \textbf{6 dB} \\
\midrule
\multirow{2}{*}{BCH $(127,113)$}
 & Derived dedicated mask ($\tau=0.5$) & 4.68 & 6.45 & 9.11 \\
 & Reused library mask ($\tau=0.3$)    & 4.66 & 6.43 & 8.89 \\
\midrule
\multirow{2}{*}{BCH $(63,39)$}
 & Derived dedicated mask ($\tau=0.5$) & 4.33 & 5.89 & 7.77 \\
 & Reused library mask ($\tau=0.3$)    & 4.17 & 5.26 & 7.37 \\
\midrule
\multirow{2}{*}{LDPC $(128,64)$}
 & Derived dedicated mask ($\tau=0.5$) & 4.35 & 6.38 & 9.44 \\
 & Reused library mask ($\tau=0.3$)    & 4.36 & 6.35 & 9.19 \\
\bottomrule
\end{tabular}
\end{table}

\section{Additional Results on Longer 5G NR LDPC Codes}
\label{app:long_nr_ldpc}

\begin{figure}[h]
\begin{center}
\includegraphics[width=0.5\textwidth]{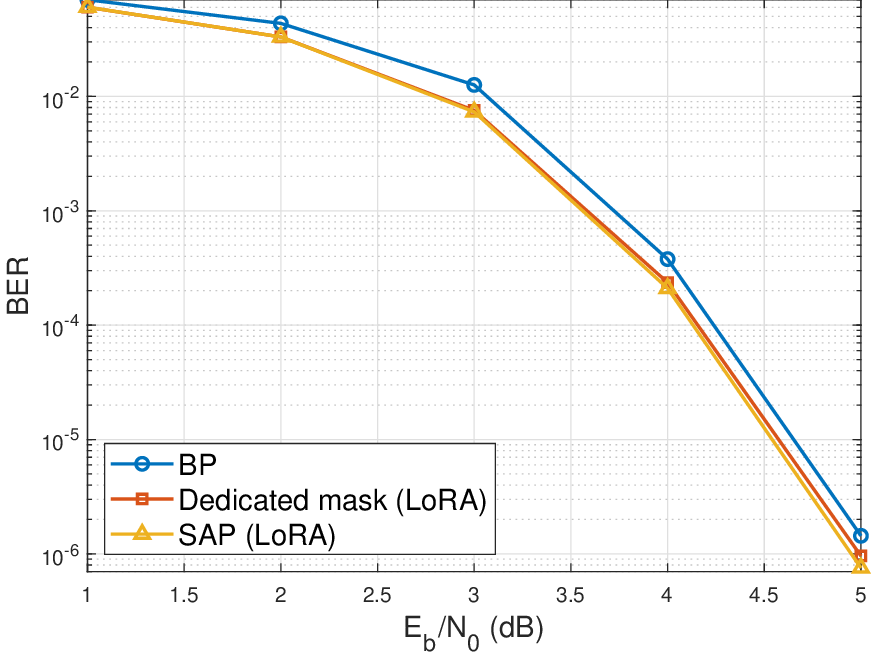}
\caption{
Decoded BER on the longer 5G NR LDPC \((512,474)\) code. We compare the dedicated per-code pruning with recovery, and SAP mask reuse with recovery.
}
\label{fig:ldpc512474}
\end{center}
\vskip -0.2in
\end{figure}

To further examine SAP on longer practical codes, we evaluate an additional 5G NR LDPC code with parameters LDPC \((512,474)\).
This code has a longer block length than the 5G NR LDPC codes reported in Section~\ref{app:long_code}, and therefore provides an additional test case for pruning-mask reuse beyond the main evaluation setting.

Figure~\ref{fig:ldpc512474} compares the decoded BER curves of the original FECCT after full fine-tuning, the dedicated pruned model with recovery, and the SAP-selected pruned model with recovery.
The SAP curve closely follows the dedicated-pruning curve across the tested \(E_b/N_0\) range, indicating that the pruning mask retrieved by SAP remains effective even for this longer LDPC code.
This result provides additional evidence that spectral retrieval can support pruning-mask reuse for longer 5G NR LDPC codes without introducing noticeable decoding-performance degradation.

\section{Generalization to Novel Code Families}
\label{app:novel_families}

The mask library and the FECCT pretraining set contain only BCH, LDPC, and polar codes.
To examine whether SAP can retrieve and reuse compatible masks for code families absent from both, we evaluate Turbo $(132,40)$ and Reed-Muller (RM) $(128,64)$ codes, whose parity-check matrices are taken from the same source as in FECCT~\citep{Choukroun2024foundation}.
For both codes, SAP retrieves the mask of polar $(128,86)$ from the library, with spectral similarities $\kappa^\star=0.82$ for Turbo and $\kappa^\star=0.71$ for RM, both above the reuse threshold $\tau=0.5$.
We then apply the same LoRA recovery as in the main experiments and compare the result with a dedicated mask derived for each code.

\begin{table}[h]
\caption{Decoding performance on code families absent from the mask library.
We report $-\ln(\mathrm{BER})$ at $E_b/N_0\in\{4,5,6\}$~\si[]{\decibel} (higher is better).
Parentheses indicate the gap $\Delta=$ SAP $-$ Dedicated.
SAP reuses the mask of polar $(128,86)$ for both codes.}
\label{tab:novel_families}
\vskip 0.15in
\centering
\small
\setlength{\tabcolsep}{6pt}
\begin{tabular}{llccc}
\toprule
\textbf{Code} & \textbf{Mask} & \textbf{4 dB} & \textbf{5 dB} & \textbf{6 dB} \\
\midrule
\multirow{2}{*}{Turbo $(132,40)$}
 & Dedicated & 3.64 & 4.87 & 6.64 \\
 & SAP & 3.60 (\textbf{$-0.04$}) & 4.81 (\textbf{$-0.06$}) & 6.51 (\textbf{$-0.13$}) \\
\midrule
\multirow{2}{*}{RM $(128,64)$}
 & Dedicated & 5.39 & 7.94 & 11.08 \\
 & SAP & 5.38 (\textbf{$-0.01$}) & 7.84 (\textbf{$-0.10$}) & 11.00 (\textbf{$-0.08$}) \\
\bottomrule
\end{tabular}
\end{table}

Table~\ref{tab:novel_families} shows that SAP stays close to the dedicated masks for both codes.
The gaps lie between $-0.13$ and $-0.01$ in $-\ln(\mathrm{BER})$, which is of the same magnitude as the gaps observed for seen and unseen codes in Table~\ref{tab:decoding_performance}.
These results suggest that SAP can retrieve and reuse a compatible mask even for code families absent from the library.

\section{Frame Error Rate (FER): Frame-Level Reliability under SAP}
\label{app:fer_ablation}

\begin{figure}[h]
\begin{center}
\begin{subfigure}[b]{0.32\textwidth}
    \centering
    \includegraphics[width=\textwidth]{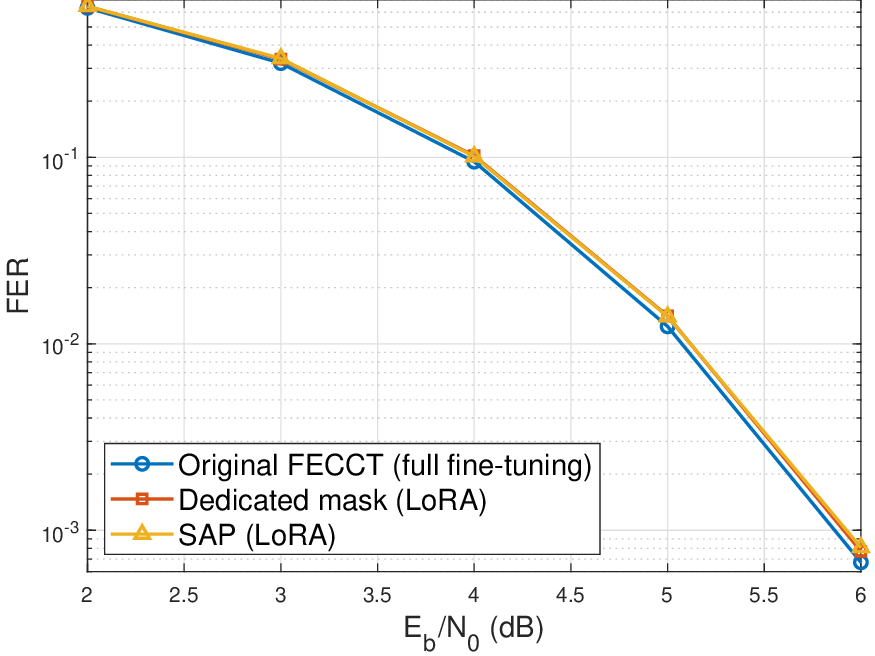}
    \caption{BCH $(63,45)$}
\end{subfigure}
\hfill
\begin{subfigure}[b]{0.32\textwidth}
    \centering
    \includegraphics[width=\textwidth]{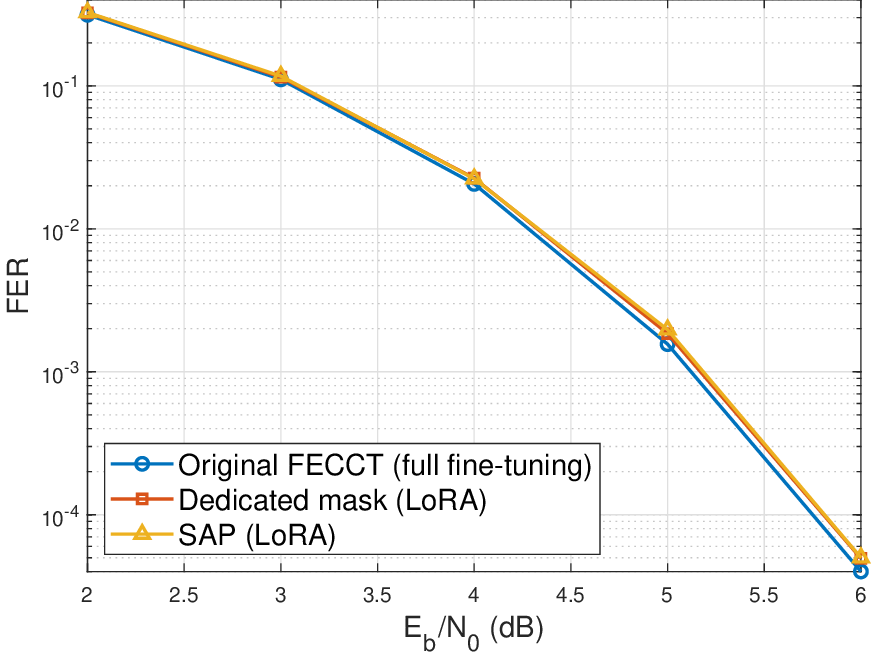}
    \caption{LDPC $(49,24)$}
\end{subfigure}
\hfill
\begin{subfigure}[b]{0.32\textwidth}
    \centering
    \includegraphics[width=\textwidth]{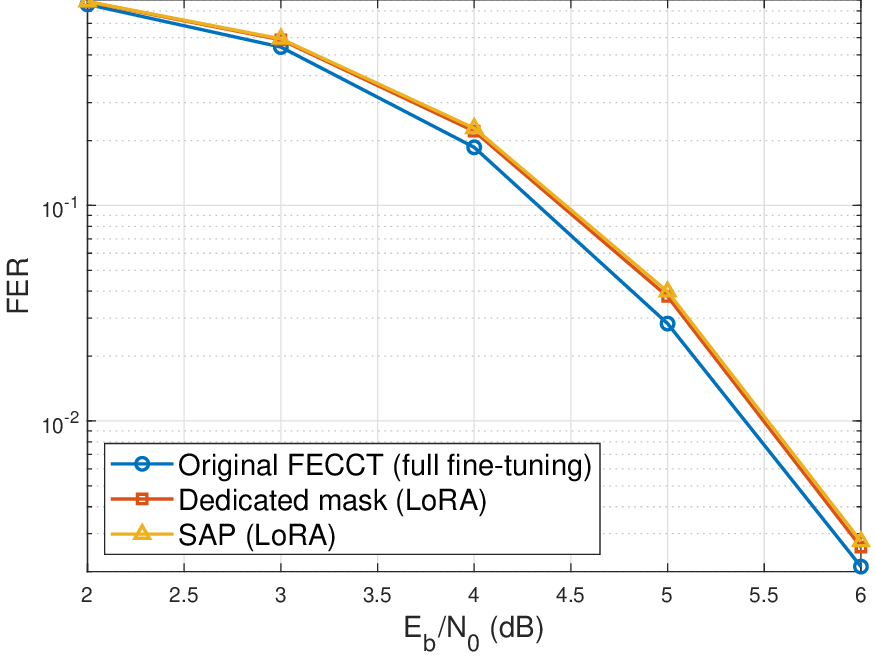}
    \caption{Polar $(128,64)$}
\end{subfigure}

\caption{
FER comparison of the pretrained FECCT baseline and the pruned decoders after recovery: dedicated per-code pruning vs.\ SAP mask reuse.
Across all three codes, SAP closely tracks the dedicated-pruning curve over the tested $E_b/N_0$ range, suggesting that spectrum-guided mask reuse preserves frame-level reliability.
}
\label{fig:fer_ablation}
\end{center}
\vskip -0.2in
\end{figure}

While the main text focuses on BER, we additionally report FER to evaluate frame-level reliability.
As shown in Figure ~\ref{fig:fer_ablation}, the SAP-reused mask achieves FER comparable to dedicated per-code pruning after recovery for BCH $(63,45)$, LDPC $(49,24)$ and polar $(128,64)$ codes.
This indicates that spectrum-guided mask reuse maintains reliability not only at the bit level (BER) but also at the frame level (FER).

\section{LoRA Rank Sensitivity: Recovery Performance vs.\ Adaptation Cost}
\label{app:ablation_lora}

\begin{table}[h]
\caption{Effect of the LoRA rank $r$ on post-pruning recovery.
We report the number of trainable parameters and $-\ln(\mathrm{BER})$ at $E_b/N_0\in\{4,5,6\}$\,dB (higher is better).
}
\label{tab:ablation_lora_params}
\vskip 0.15in
\centering
\small
\setlength{\tabcolsep}{6pt}
\begin{tabular}{l c r ccc}
\toprule
\textbf{Code} & \textbf{LoRA $r$} & \textbf{Trainable} & \textbf{4 dB} & \textbf{5 dB} & \textbf{6 dB} \\
\midrule
\multirow{3}{*}{Polar $(64,32)$}
 & 4   & 44{,}340  & 5.98 & 8.11 & 10.68 \\
 & 8  & 88{,}200  & 6.16 & 8.35 & 11.48 \\
 & 16 & 177{,}360 & 6.21 & 8.49 & 11.50 \\
\midrule
\multirow{3}{*}{LDPC $(121,70)$}
 & 4   & 43{,}468  & 6.38 & 10.06 & 15.38 \\
 & 8  & 87{,}176  & 6.43 & 10.12 & 15.65 \\
 & 16 & 173{,}872 & 6.45 & 10.32 & 15.76 \\
\bottomrule
\end{tabular}
\end{table}

Table~\ref{tab:ablation_lora_params} studies how the LoRA rank $r$ trades off recovery quality against adaptation cost.
As expected, increasing $r$ increases the number of trainable parameters approximately linearly.
For polar $(64,32)$, moving from $r=4$ to $8$ roughly doubles trainable parameters (44.3k $\rightarrow$ 88.2k) and yields a clear recovery gain, whereas increasing to $r=16$ again doubles the parameters (88.2k $\rightarrow$ 177.4k) but provides only marginal improvement.
A similar diminishing-return trend is observed for LDPC $(121,70)$: $r=4 \rightarrow 8$ improves recovery with a near-doubling of parameters (43.5k $\rightarrow$ 87.2k), while $r=8 \rightarrow 16$ (87.2k $\rightarrow$ 173.9k) yields comparatively smaller gains.
Based on this diminishing-return behavior, we use $r = 8$ as the default recovery configuration, as it achieves near-saturated recovery performance with substantially fewer trainable parameters than higher-rank settings.

\section{Effect of Knowledge Distillation During Recovery}
\label{app:ablation_kd}

\begin{table}[h]
\caption{Impact of the KD loss during recovery on two representative codes.
We report $-\ln(\mathrm{BER})$ at $E_b/N_0\in\{4,5,6\}$~\si[]{\decibel} (larger is better).
BCE $+$ KD uses the default setting ($\gamma=1$) with the unpruned model as teacher, whereas BCE disables KD ($\gamma=0$) and optimizes only the BCE term.}
\label{tab:ablation_kd_two_codes}
\vskip 0.15in
\centering
\small
\setlength{\tabcolsep}{6pt}
\begin{tabular}{llccc}
\toprule
\textbf{Code} & \textbf{Recovery setting} & \textbf{4 dB} & \textbf{5 dB} & \textbf{6 dB} \\
\midrule
\multirow{2}{*}{LDPC $(121,70)$} 
 & BCE $+$ KD ($\gamma=1$)  & 6.43 & 10.12 & 15.65 \\
 & BCE ($\gamma=0$) & 6.37 & 10.04 & 15.48 \\
\midrule
\multirow{2}{*}{Polar $(64,32)$} 
 & BCE $+$ KD ($\gamma=1$)  & 6.16 & 8.35 & 11.48 \\
 & BCE ($\gamma=0$) & 6.11 & 8.24 & 11.19 \\
\bottomrule
\end{tabular}
\end{table}

Table~\ref{tab:ablation_kd_two_codes} indicates that disabling KD leads to a small but consistent degradation in recovery performance under the same retraining budget.
Across both codes and all evaluated SNRs, using both the BCE loss and the KD loss improves BER compared to using BCE loss alone.
These results support using KD as an auxiliary signal for short post-pruning recovery when a reliable teacher (the unpruned model) is available.

\section{Full Fine-Tuning Results of the Original (Unpruned) FECCT}
\label{app:full_finetune_results}

\begin{figure*}[h]
\vskip 0.10in
\centering

\begin{subfigure}[b]{0.32\textwidth}
    \centering
    \includegraphics[width=\textwidth]{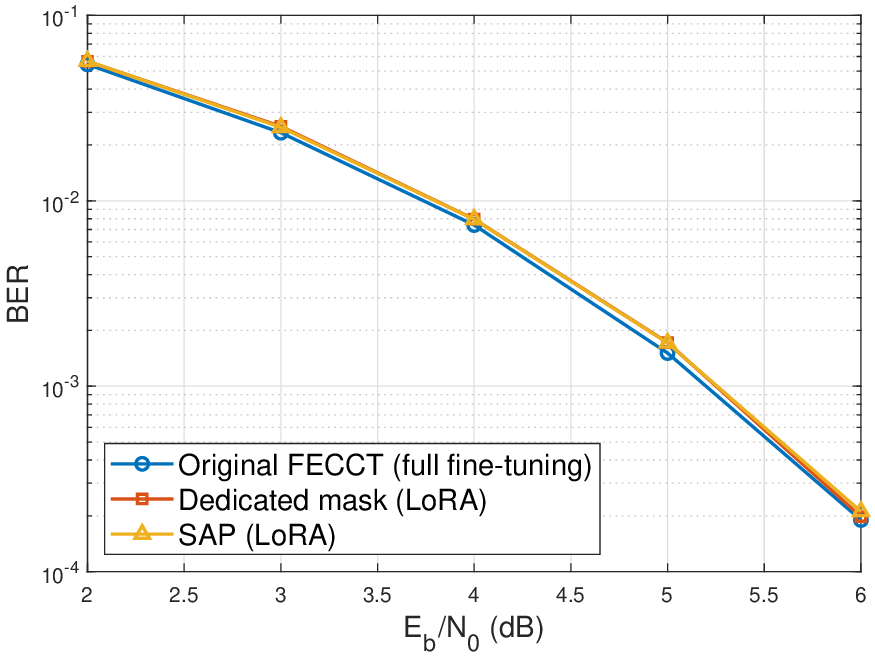}
    \caption{BCH $(31,21)$}
    \label{fig:ft_bch_31_11}
\end{subfigure}
\hfill
\begin{subfigure}[b]{0.32\textwidth}
    \centering
    \includegraphics[width=\textwidth]{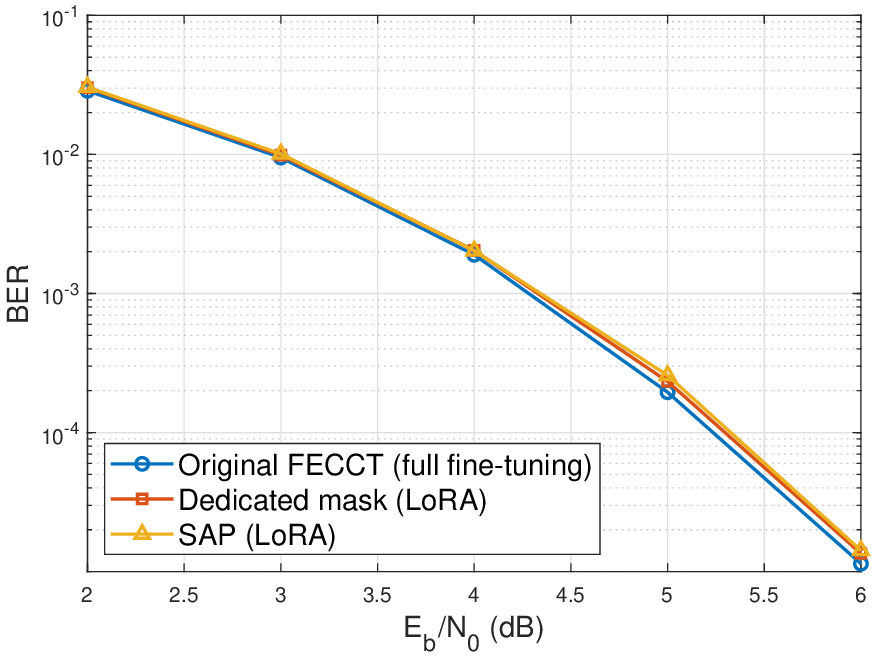}
    \caption{BCH $(31,11)$}
    \label{fig:ft_bch_31_21}
\end{subfigure}
\hfill
\begin{subfigure}[b]{0.32\textwidth}
    \centering
    \includegraphics[width=\textwidth]{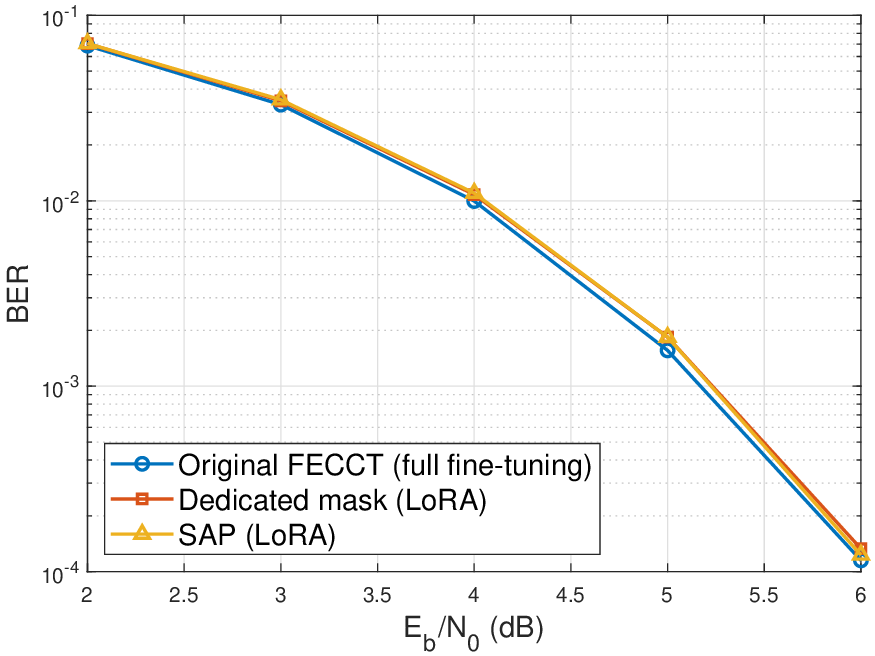}
    \caption{BCH $(63,36)$}
    \label{fig:ft_bch_63_36}
\end{subfigure}

\vskip 0.08in

\begin{subfigure}[b]{0.32\textwidth}
    \centering
    \includegraphics[width=\textwidth]{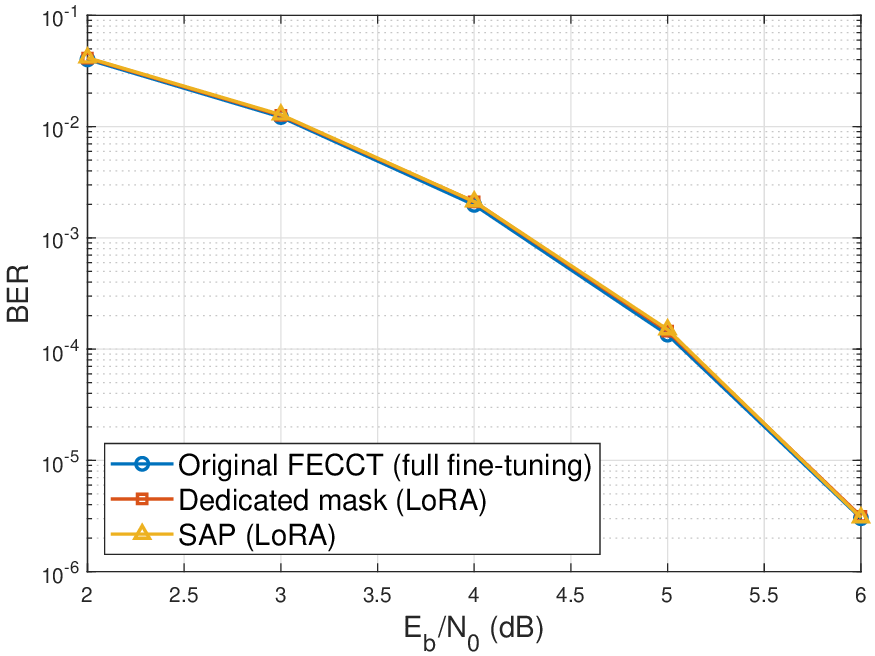}
    \caption{LDPC $(49,24)$}
    \label{fig:ft_ldpc_49_24}
\end{subfigure}
\hfill
\begin{subfigure}[b]{0.32\textwidth}
    \centering
    \includegraphics[width=\textwidth]{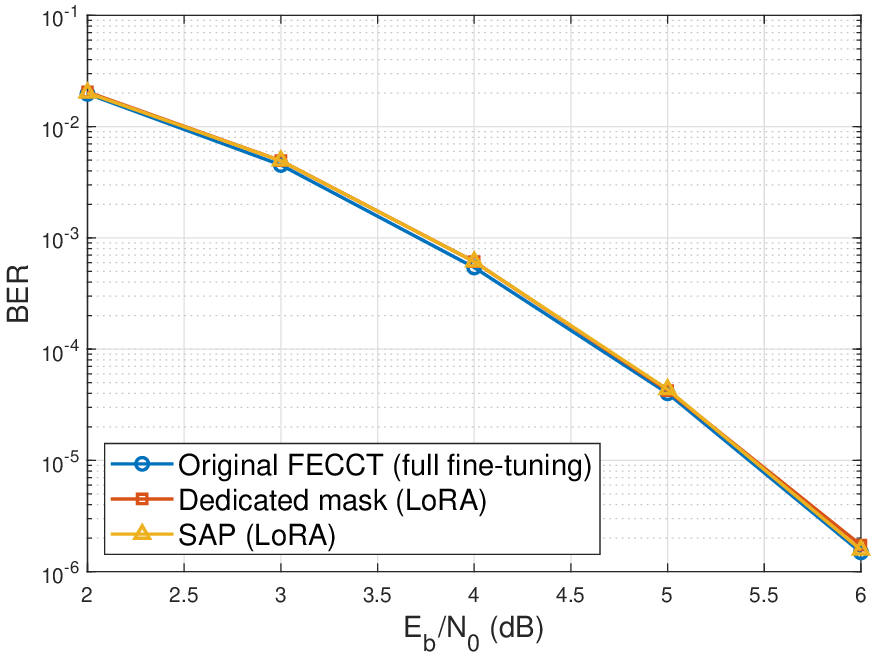}
    \caption{LDPC $(64,48)$}
    \label{fig:ft_ldpc_64_48}
\end{subfigure}
\hfill
\begin{subfigure}[b]{0.32\textwidth}
    \centering
    \includegraphics[width=\textwidth]{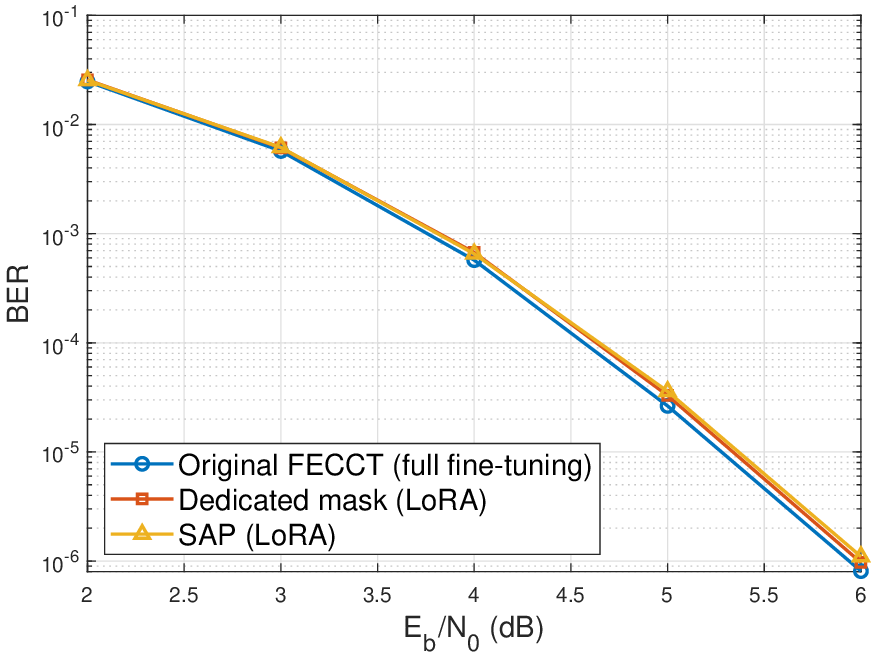}
    \caption{LDPC $(96,72)$}
    \label{fig:ft_ldpc_96_72}
\end{subfigure}

\vskip 0.08in

\begin{subfigure}[b]{0.32\textwidth}
    \centering
    \includegraphics[width=\textwidth]{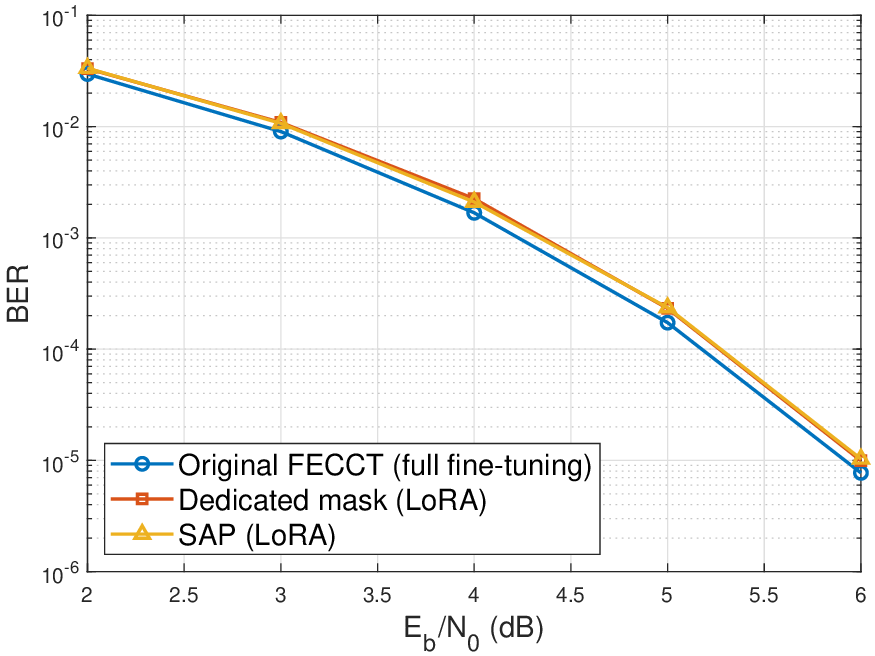}
    \caption{Polar $(64,32)$}
    \label{fig:ft_polar_64_32}
\end{subfigure}
\hfill
\begin{subfigure}[b]{0.32\textwidth}
    \centering
    \includegraphics[width=\textwidth]{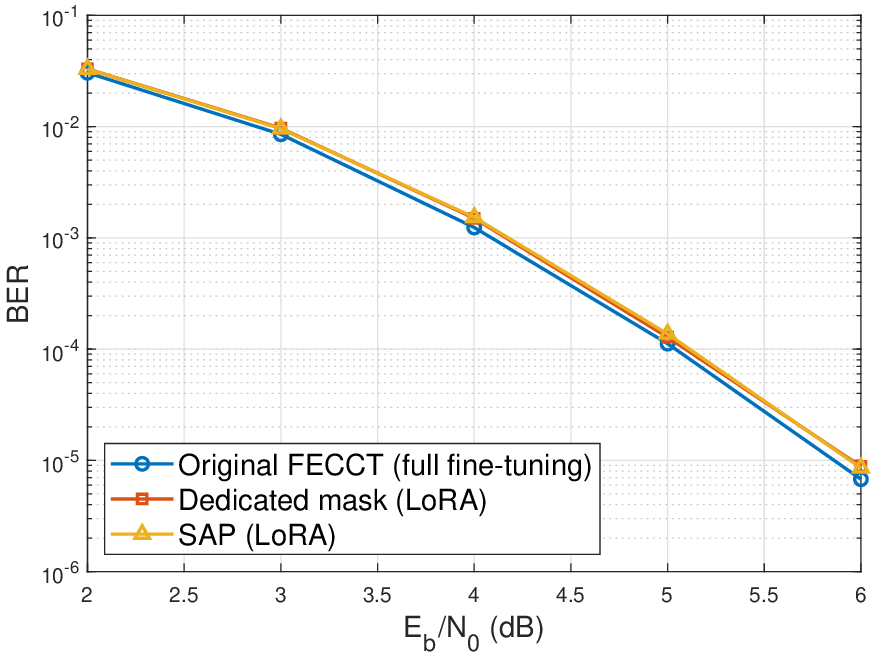}
    \caption{Polar $(64,43)$}
    \label{fig:ft_polar_64_43}
\end{subfigure}
\hfill
\begin{subfigure}[b]{0.32\textwidth}
    \centering
    \includegraphics[width=\textwidth]{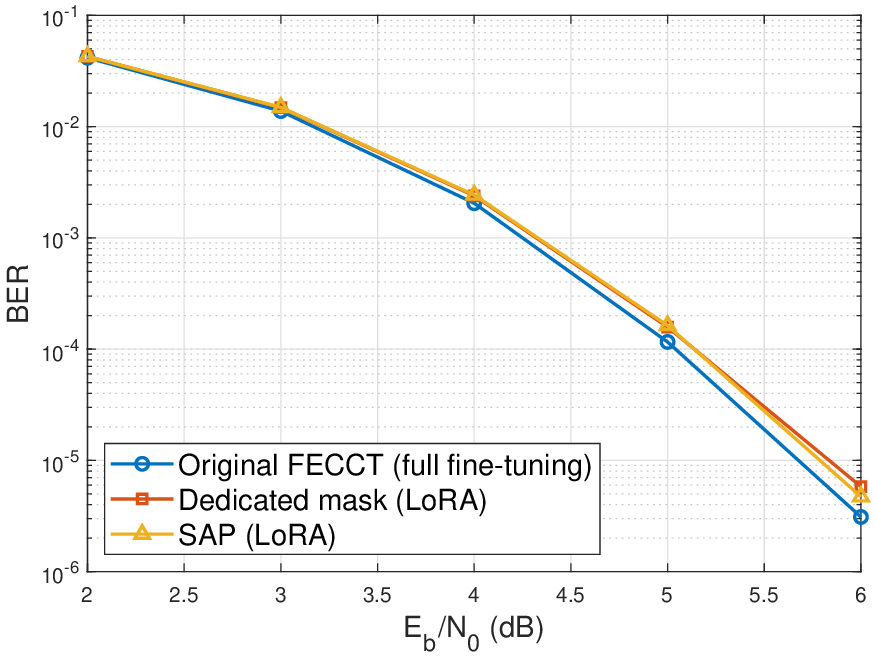}
    \caption{Polar $(128,96)$}
    \label{fig:ft_polar_128_96}
\end{subfigure}
\caption{Decoded BER comparison of the pretrained FECCT baseline, the dedicated pruned model, and the SAP model.
Results are shown for the codes listed in Table~\ref{tab:decoding_performance}.
\label{fig:full_finetune_grid}
}
\vskip -0.15in
\end{figure*}

For completeness, Figure~\ref{fig:full_finetune_grid} shows the full fine-tuning results of the original (unpruned) FECCT on the codes listed in Table~\ref{tab:decoding_performance}.
After full fine-tuning, across most codes, the unpruned FECCT yields only a marginal SNR gain relative to pruned models, typically within $\le$ \SI{0.1}{\decibel} over the tested $E_b/N_0$ range.

\section{Pruning-Ratio Sensitivity}
\label{app:pruning_ratio}

\begin{figure*}[h]
\vskip 0.10in
\centering
\begin{subfigure}[b]{0.32\textwidth}
    \centering
    \includegraphics[width=\textwidth]{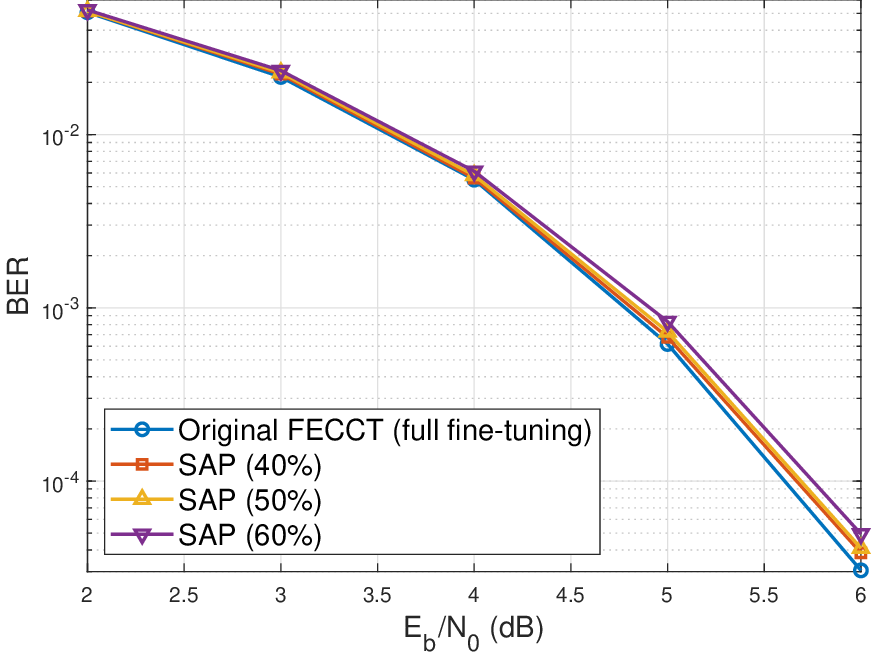}
    \caption{BCH $(63,45)$}
\end{subfigure}
\hfill
\begin{subfigure}[b]{0.32\textwidth}
    \centering
    \includegraphics[width=\textwidth]{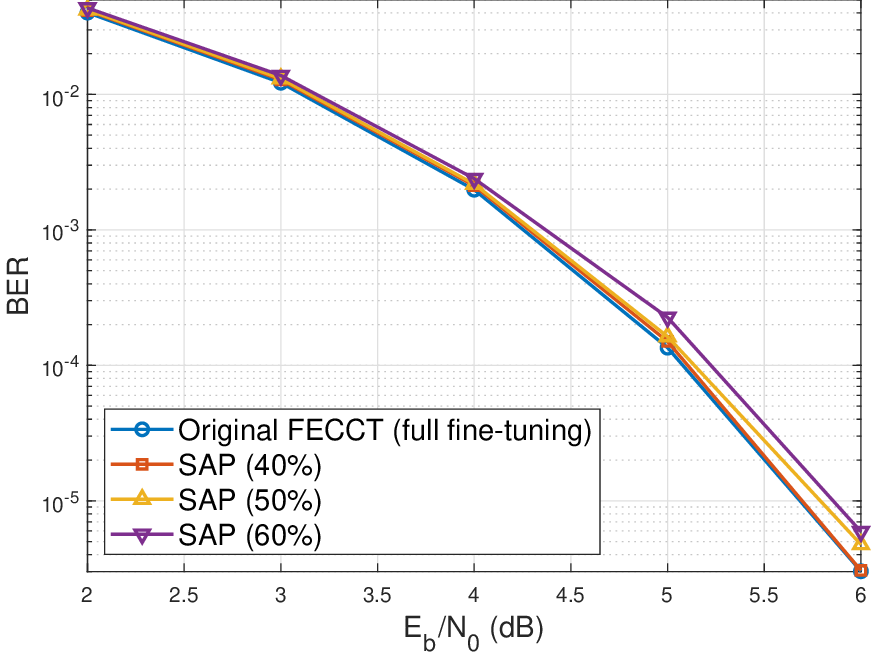}
    \caption{LDPC $(49,24)$}
\end{subfigure}
\hfill
\begin{subfigure}[b]{0.32\textwidth}
    \centering
    \includegraphics[width=\textwidth]{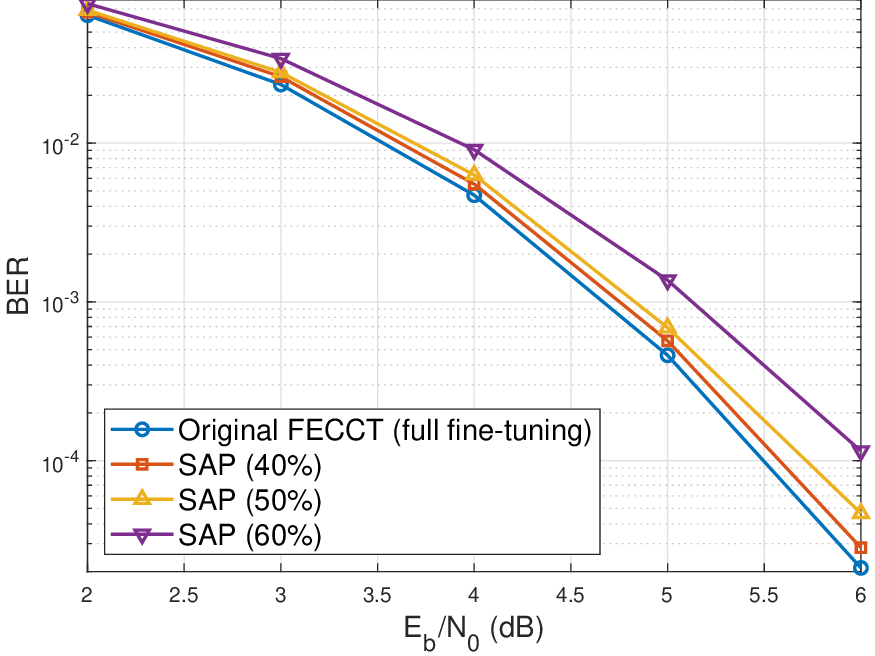}
    \caption{Polar $(128,64)$}
\end{subfigure}

\caption{
Decoded BER under varying pruning ratios.
}
\label{fig:pruning_ratio}
\vskip -0.15in
\end{figure*}

Figure~\ref{fig:pruning_ratio} reports BER curves under several pruning ratios for three representative codes, BCH $(63,45)$, LDPC $(49,24)$, and polar $(128,64)$ codes.
In the main experiments, we adopt a pruning ratio of \SI{40}{\percent} across code families to ensure a consistent efficiency target.
We observe that the maximum tolerable pruning ratio can be code-dependent: some codes admit more aggressive structured pruning with minimal degradation, whereas others show earlier performance loss as the pruning ratio increases.
For this reason, we adopt \SI{40}{\percent} as a conservative default that remains broadly stable across the evaluated code families, while still providing meaningful efficiency gains.

\section{Monte Carlo Evaluation and Confidence Intervals}
\label{app:monte_carlo_eval}

Following standard evaluation practice for neural ECC decoders, BER and FER are estimated by Monte Carlo sampling over independent AWGN noise realizations.
For each code, decoder, and \(E_b/N_0\) value, noisy received vectors are generated as
\[
y = \mathrm{BPSK}(x) + z,
\qquad
z \sim \mathcal{N}(0,\sigma^2 I),
\]
where \(\sigma\) is determined by the target \(E_b/N_0\).
The same evaluation protocol is used for all compared methods, including the original FECCT baseline, dedicated per-code pruning, and SAP mask reuse.

Let \(N_f\) denote the number of evaluated frames and \(n\) the block length.
Let \(E_f\) be the number of frame errors and \(E_b\) be the number of bit errors over \(N_b=N_f n\) decoded bits.
We estimate
\[
\widehat{\mathrm{FER}}=\frac{E_f}{N_f},
\qquad
\widehat{\mathrm{BER}}=\frac{E_b}{N_f n}.
\]
In our evaluation, each test point is simulated using at least \(10^5\) frames and sampling continues until more than 100 frame errors are observed, unless the maximum simulation budget is reached.
Thus, the reported BER and FER values are Monte Carlo estimates of the corresponding decoding error probabilities under the AWGN channel.

To quantify the statistical uncertainty of these Monte Carlo estimates, we treat BER and FER as binomial proportion estimates.
For a generic error-rate estimate \(\hat p=E/N\), we compute the Wilson 95\% confidence interval as
\[
\mathrm{CI}_{95\%}(\hat p)
=
\frac{
\hat p + \frac{z^2}{2N}
\pm
z\sqrt{\frac{\hat p(1-\hat p)}{N}+\frac{z^2}{4N^2}}
}{
1+\frac{z^2}{N}
},
\qquad z=1.96.
\]
For FER, we use \((E,N)=(E_f,N_f)\), and for BER, we use \((E,N)=(E_b,N_f n)\).

As an example, for LDPC \((49,24)\), the evaluated number of frames is
\[
N_f=[100352,100352,100352,100352,2017280]
\]
for \(E_b/N_0=[2,3,4,5,6]\) dB, respectively.
At \(E_b/N_0=6\) dB, this corresponds to \(N_b=98{,}846{,}720\) decoded bits.
The observed \(\mathrm{FER}=5.01\times10^{-5}\) corresponds to approximately \(101\) frame errors, and the observed \(\mathrm{BER}=3.06\times10^{-6}\) corresponds to approximately \(302\) bit errors.
The corresponding Wilson 95\% confidence intervals are approximately
\[
\mathrm{FER}\in[4.12,6.09]\times10^{-5},
\qquad
\mathrm{BER}\in[2.73,3.42]\times10^{-6}.
\]

\section{Limitations}\label{app:limitations}

SAP is developed and evaluated on FECCT~\citep{Choukroun2024foundation}, a representative foundation-style transformer ECC decoder that shares a single parameter set across heterogeneous code families and code parameters. 
Another foundation-style transformer decoder for ECC has recently been presented~\citep{Park2026Efficient}, and applying SAP to such architectures is left for future work. 
We note that the spectral signature $\phi(H) = [\lambda_1, \lambda_2]$ is computed entirely from the parity-check matrix and is independent of the specific transformer architecture, so we expect SAP's retrieval mechanism to extend to other foundation-style decoders that share parameters across codes; however, this generalization is not empirically verified in this work.


\newpage
\input{checklist.tex}

\end{document}

%% file: checklist.tex
\section*{NeurIPS Paper Checklist}
\begin{enumerate}

\item {\bf Claims}
    \item[] Question: Do the main claims made in the abstract and introduction accurately reflect the paper's contributions and scope?
    \item[] Answer: \answerYes{} 
    \item[] Justification: The abstract and introduction state the proposed SAP framework, its spectral retrieval criterion, LoRA-based recovery, and the evaluated efficiency and decoding-performance claims, which are supported by the experiments in Sections~\ref{sec:experiments} and~\ref{sec:analysis}.

\item {\bf Limitations}
    \item[] Question: Does the paper discuss the limitations of the work performed by the authors?
    \item[] Answer: \answerYes{} 
    \item[] Justification: We mention our limitations in Appendix~\ref{app:limitations}.

\item {\bf Theory assumptions and proofs}
    \item[] Question: For each theoretical result, does the paper provide the full set of assumptions and a complete (and correct) proof?
    \item[] Answer: \answerNA{} 
    \item[] Justification: The paper does not introduce new theoretical results requiring original proofs; graph-spectral and coding-theoretic properties used to motivate SAP are stated with references in Section~\ref{sec:spectrum} and Appendix~\ref{app:spectrum_ecc}.

    \item {\bf Experimental result reproducibility}
    \item[] Question: Does the paper fully disclose all the information needed to reproduce the main experimental results of the paper to the extent that it affects the main claims and/or conclusions of the paper (regardless of whether the code and data are provided or not)?
    \item[] Answer: \answerYes{} 
    \item[] Justification: The paper provides the SAP algorithm in Appendix~\ref{app:MS_algorithm}, describes the experimental setup in Section~\ref{sec:experiments}, and lists the full implementation details in Appendix~\ref{app:implementation_details} and the pretraining code set in Appendix~\ref{app:pretrain_codes}.

\item {\bf Open access to data and code}
    \item[] Question: Does the paper provide open access to the data and code, with sufficient instructions to faithfully reproduce the main experimental results, as described in supplemental material?
    \item[] Answer: \answerYes 
    \item[] Justification: The code is publicly available at https://github.com/iil-postech/sap. The repository includes the pretrained FECCT checkpoint and scripts for pruning-mask library construction, spectral retrieval, and LoRA retraining, and its README gives the installation instructions, commands, and arguments.

\item {\bf Experimental setting/details}
    \item[] Question: Does the paper specify all the training and test details (e.g., data splits, hyperparameters, how they were chosen, type of optimizer) necessary to understand the results?
    \item[] Answer: \answerYes{} 
    \item[] Justification: The paper specifies the main experimental settings in Section~\ref{sec:experiments} and Appendix~\ref{app:implementation_details}.

\item {\bf Experiment statistical significance} 
    \item[] Question: Does the paper report error bars suitably and correctly defined or other appropriate information about the statistical significance of the experiments?
    \item[] Answer: \answerYes{}
    \item[] Justification: BER and FER are estimated through Monte Carlo sampling over independent AWGN noise realizations, and Appendix~\ref{app:monte_carlo_eval} describes the sample sizes, error counts, and Wilson confidence intervals used to quantify statistical uncertainty.

\item {\bf Experiments compute resources}
    \item[] Question: For each experiment, does the paper provide sufficient information on the computer resources (type of compute workers, memory, time of execution) needed to reproduce the experiments?
    \item[] Answer: \answerYes{} 
    \item[] Justification: We provide information about the computer resources used in Section~\ref{sec:experiments}, and report wall-clock costs for pruning-mask derivation and spectral reuse in Appendix~\ref{app:mask_cost}.
    
\item {\bf Code of ethics}
    \item[] Question: Does the research conducted in the paper conform, in every respect, with the NeurIPS Code of Ethics \url{https://neurips.cc/public/EthicsGuidelines}?
    \item[] Answer: \answerYes{} 
    \item[] Justification: We have read and understood the code of ethics and have made every effort to adhere to it.

\item {\bf Broader impacts}
    \item[] Question: Does the paper discuss both potential positive societal impacts and negative societal impacts of the work performed?
    \item[] Answer: \answerNA{} 
    \item[] Justification: This work focuses on improving the efficiency of transformer-based error-correcting code decoders in standard communication simulations and does not involve human subjects, private data, generative content, or deployment in decision-making systems with direct societal impact.
    
\item {\bf Safeguards}
    \item[] Question: Does the paper describe safeguards that have been put in place for responsible release of data or models that have a high risk for misuse (e.g., pre-trained language models, image generators, or scraped datasets)?
    \item[] Answer: \answerNA{} 
    \item[] Justification: The paper does not release high-risk models, scraped datasets, generative models, or data with misuse potential; it uses standard ECC simulation settings and proposes an efficiency method for transformer-based decoders.

\item {\bf Licenses for existing assets}
    \item[] Question: Are the creators or original owners of assets (e.g., code, data, models), used in the paper, properly credited and are the license and terms of use explicitly mentioned and properly respected?
    \item[] Answer: \answerYes{} 
    \item[] Justification: The paper properly cites the existing methods and code families used for comparison and evaluation, including ECCT/FECCT; no scraped datasets or third-party assets with additional licensing constraints are introduced.

\item {\bf New assets}
    \item[] Question: Are new assets introduced in the paper well documented and is the documentation provided alongside the assets?
    \item[] Answer: \answerNA{} 
    \item[] Justification: The paper does not introduce or release new datasets, models, or other assets at submission time.

\item {\bf Crowdsourcing and research with human subjects}
    \item[] Question: For crowdsourcing experiments and research with human subjects, does the paper include the full text of instructions given to participants and screenshots, if applicable, as well as details about compensation (if any)? 
    \item[] Answer: \answerNA{} 
    \item[] Justification: This work does not use crowdsourcing or human subjects.

\item {\bf Institutional review board (IRB) approvals or equivalent for research with human subjects}
    \item[] Question: Does the paper describe potential risks incurred by study participants, whether such risks were disclosed to the subjects, and whether Institutional Review Board (IRB) approvals (or an equivalent approval/review based on the requirements of your country or institution) were obtained?
    \item[] Answer: \answerNA{} 
    \item[] Justification: This work does not use crowdsourcing or human subjects.

\item {\bf Declaration of LLM usage}
    \item[] Question: Does the paper describe the usage of LLMs if it is an important, original, or non-standard component of the core methods in this research? Note that if the LLM is used only for writing, editing, or formatting purposes and does \emph{not} impact the core methodology, scientific rigor, or originality of the research, declaration is not required.
    \item[] Answer: \answerNA{} 
    \item[] Justification: We used an LLM solely for editing assistance.

\end{enumerate}